\RequirePackage{fix-cm}
\documentclass[smallextended]{svjour3}       % onecolumn (second format)

\usepackage{todonotes}
\usepackage{graphicx}
\usepackage[misc]{ifsym}
\usepackage{amssymb,enumitem,booktabs,xcolor,todonotes,setspace}
\usepackage[font=small,labelfont=bf,labelsep=none]{caption}
\usepackage[top=1.4in, bottom=1.2in, left=1.35in, right=1.35in]{geometry}
\usepackage{natbib}
\usepackage{tikz}
\usepackage{url} % not crucial - just used below for the URL
\usepackage[pdftex,colorlinks=true]{hyperref}
\definecolor{darkblue}{rgb}{0,0,.6}
\hypersetup{citecolor=darkblue,linkcolor=darkblue,urlcolor=darkblue}
\usepackage{amsmath}
\usepackage{bm}
\usepackage{epsfig}
\usepackage{graphics}
\usepackage{lscape}
\usepackage{rotating}
\usepackage{indentfirst}
\usepackage{booktabs}
\usepackage{multirow}
\usepackage{float}
\usepackage{lipsum}
\usepackage{subfigure}
\renewcommand{\figurename}{Fig.}
\providecommand{\U}[1]{\protect\rule{.1in}{.1in}}

\graphicspath{{plots/}}
\journalname{Statistical Papers}

\definecolor{lime}{HTML}{A6CE39}
\DeclareRobustCommand{\orcidicon}{
\begin{tikzpicture}
\draw[lime, fill=lime] (0,0)
circle[radius=0.16]
node[white]{{\fontfamily{qag}\selectfont \tiny \.{I}D}};
\end{tikzpicture}
\hspace{-2mm}
}
\foreach \x in {A, ..., Z}{%
\expandafter\xdef\csname orcid\x\endcsname{\noexpand\href{https://orcid.org/\csname orcidauthor\x\endcsname}{\noexpand\orcidicon}}
}

\begin{document}

\title{Change Point Detection for Nonparametric Regression under Strongly Mixing Process
}
\author{Qing Yang \hspace{-1.5mm}\orcidA{}
        \and Yu-Ning Li \hspace{-1.5mm}\orcidB{}
        \and Yi Zhang \hspace{-1.5mm}\orcidC{}
}
\institute{Qing Yang(\Letter) \at
              School of Mathematical Sciences, Zhejiang University, Hangzhou 310027, China.\\
              \email{yq1995@zju.edu.cn}
           \and
           Yu-Ning Li \at
             School of Mathematical Sciences, Zhejiang University, Hangzhou 310027, China. \\
             %\emph{Present address: Department of Economics and Related Studies, University of York, Heslington, YO10 5DD, UK}\\
             \email{mark\_li@foxmail.com}
           \and
            Yi Zhang \at
              School of Mathematical Sciences, Zhejiang University, Hangzhou 310027, China. \\
              \email{zhangyi63@zju.edu.cn}
}
\date{Received: date / Accepted: date}

\maketitle

\begin{abstract}
In this article, we consider the estimation of the structural change point in the nonparametric model with dependent observations. We introduce a maximum-CUSUM-estimation procedure, where the CUSUM statistic is constructed based on the sum-of-squares aggregation of the difference of the two Nadaraya-Watson estimates using the observations before and after a specific time point. Under some mild conditions, we prove that the statistic tends to zero almost surely if there is no change, and is larger than a threshold asymptotically almost surely otherwise, which helps us to obtain a threshold-detection strategy. Furthermore, we demonstrate the strong consistency of the change point estimator. In the simulation, we discuss the selection of the bandwidth and the threshold used in the estimation, and show the robustness of our method in the long-memory scenario. We implement our method to the data of Nasdaq 100 index and find that the relation between the realized volatility and the return exhibits several structural changes in 2007--2009.
{\subclass{Primary 62G05; Secondary 62G08\and62G20\and62M10}}
\end{abstract}

\section{Introduction}\label{intro}
\renewcommand{\theequation}{1.\arabic{equation}}
\setcounter{equation}{0}

Structural change is a variation of a system over time, which is usually unexpected but can lead to a huge estimation and prediction error when we specify a time-invariant model. To deal with this, one can divide the sample set into two sub-samples on the time interval, and model each sub-sample separately. The key issue here is how to detect the changes and separate samples. More specifically, one should judge whether the change point exists, and, if it exists, where it is located.
Change point detection methods are divided into two main branches: on-line methods, which aim to detect changes as soon as they occur in a real-time setting, and off-line methods that retrospectively detect changes when all samples are received. The former task is often referred to as event or anomaly detection, while the latter is sometimes called signal segmentation. In this paper, our research core is off-line problem with all samples already collected.
In addition, we will only focus on abrupt change, though
gradual change is another kind of structure change that has attracted extensive attention and research. Readers can refer to \cite{gradualchange02} for the hypothesis test of gradual change.

Structural change, as suggested by the name, is always related to a specific structure. When a parametric model is specified, some of the parameters in the model may change over time. More general, any feature of a structure such as mean, variance and quantile can exhibit changes.
Readers can refer to the excellent literature published recently such as \cite{mean18-1}, \cite{mean19-1} and \cite{mean19-2} for mean changes, \cite{var19} for variance changes, \cite{otherquantity18} for other quantities of interest, and \cite{distribution14} for distribution changes.
Recently, an analogous mean-change-point setup became very popular in the functional data framework and motivated the new development, see, for instance, \cite{functional1} and \cite{functional2}.

Regression functions, which describe the relationship between  regressand and  regressors, of course, can change over time.
It is of great interest for researchers to detect the structural change of regression functions, especially the linear ones.
For example, \cite{pl75} tested the instability of the regression parameters based on the CUSUM of recursive residuals and \cite{pl92} studied CUSUM tests based on OLS  residuals without standardising it.
More recently, \cite{pl16-1} used the jackknife empirical likelihood method for detecting the change in the parameters of the linear regression. \cite{pl17-1} studied the multiple change point problem with the change point numbers known. \cite{pl19-1} considered a high-dimensional case without grid search and they added a change-inducing variable in their linear regression model.
In terms of non-linear regressions, \cite{GB05} studied the structural change of the logistic regression. %\cite{sp06} under the semi-parametric model.

The most researches are devoted to the structural change of the parametric model, resulting in relatively few results and literature in the field of the structural change of the nonparametric models.
%As mentioned above, two existing relevant topics are more popular, and attract the remaining attention of the researchers to the nonparametric models.
To detect the changes in the nonparametric model, \cite{np2} proposed a CUSUM test and \cite{MN19} modified it to achieve consistency. However, they did not estimate the changes after the change points were found.
\cite{np1} studied the weak consistency of the change point estimator for a long-memory process,
which motivates our change-point estimation research. Recently, \cite{np4} proposed an estimator based on Kolmogorov-Smirnov functional of the marked empirical process of residuals.

The existing change point literature usually falls into
two main categories: the change point detection and the change point estimation. The change point detection is usually performed by the means of some
statistical test, while the change
point estimation tries to locate the change points.
We investigate both aspects of the change point problem by using a threshold-based method, which was developed by \cite{bs5}, \cite{th1} and \cite{th2}.
This method does not require a prior knowledge that
the change point is really present in the model.
With some ad-hoc threshold, consistency of the estimator is guaranteed, and no
formal statistical test is implemented to controls Type I errors.
We emphasize that the threshold method is consistent in estimating the number of change points. If there are multiple change points, though it is not investigated in this paper, the binary segmentation method can be used. Interested readers can refer to \cite{bs1}, \cite{bs2}, \cite{bs3}, \cite{bs4}, \cite{bs5}
and the references therein.

We will investigate the following nonparametric model with a structural change,
\begin{eqnarray}
Y_t=\left\{
\begin{aligned}
\varphi_1(X_t)+\epsilon_t,\quad 1\leq t \leq k,\\
\varphi_2(X_t)+\epsilon_t,\quad k<t\leq n,\\
\end{aligned}
\right.
\end{eqnarray}
where ${ Y_t}$ is the regressand and $X_t$ is the regressor, $\varphi_1(x)$ and $\varphi_2(x)$ are two different regression functions, and $k$ is an unknown change point.

It is important to make it clear that the problem we focus on is different from regression discontinuity.
Regression discontinuity  assumes a discontinuous regression function, and is prevailing in the treatment-effect analysis,
see \cite{fix92} and \cite{np3} for fixed-design setting, \cite{random04} for  random-design setting,  \cite{fix93} and \cite{fix98} for multiple discontinuity points detection problem, and \cite{CL17} for reviews.
Discontinuity points in a regression function are sometimes referred to as change points as well, so they can be ambiguous.
For instance, \cite{discontinuities} considered a nonparametric regression with $\alpha$-mixing dependence, which seems similar to this paper, but they
investigated the discontinuities, which is fundamentally different.
\cite{DH00} tried to estimate the change points and discontinuity points in a uniform framework by estimating the right and left limits of an extended regression function including the scaling time as an extra variable.
This method is very common  in regression discontinuity, but does not take into account the priori information that the regression function is indeed not changed in two subsamples. In other words, the extended regression function is flat in the direction of time variable.
Therefore the locally comparison method is intuitively not optimal which only uses limited data and information.
In comparison, our method still identifies two regression functions before and after the change point rather than one discontinuity but time-invariant regression function.
To estimate the change point, two regression functions are estimated using subsamples and compared across the entire domain of the covariate.

We make the following contributions. Firstly,
we construct a CUSUM statistic to detect the structural change in the nonparametric regression model. The CUSUM statistic is aggregated using the sum-of-squares, which does not focus on only one extremal point like \cite{np1}.
Secondly, we not only establish the consistency of the change point estimator but also derive an asymptotic upper bound of the CUSUM statistic when there is no change point and an asymptotic lower bound when there is a change point. This result leads to a threshold method that can judge whether there is a change point.
We show in the simulation that our method makes few false positive and false negative determinations.
Last but not least,
we show a surprising result that the CUSUM statistic constructed by the Nadaraya-Watson (N-W) estimator performs better than that constructed by the local linear estimator, because the N-W estimator is more sensitive to the observations or outliers which do not belong to the same stable period.

In Section \ref{sec2}, we introduce the basic preliminary of the nonparametric model and propose a sum-of-squares aggregated CUSUM statistic to locate the position of change point.
In Section \ref{sec3}, we propose a threshold method based on the CUSUM statistic to detect whether the regression function changes over time.
Under strongly mixing assumption, some asymptotic results are established. In Section \ref{sec4}, we do simulations and show how the bandwidth and the threshold in the estimation can be selected in practice. We show that our method has a superb performance in comparison with \cite{np1}'s method and the method of the local linear estimation. Section \ref{sec5} is an application of our method to real data. Section \ref{sec6} concludes. To keep fluency, we relegate all proofs in the appendix.

\bigskip

For convenience, we unify the notations in the rest part of this paper as below:
\begin{description}
\item {\textbf{{1}}}.\ {\sl Denote the largest integer not greater than $a$ as $\lfloor a\rfloor$.}
\item {\textbf{{2}}}.\ {\sl Let $\pi\bigotimes \psi$ denote a bivariate function $\pi(x_1)\cdot \psi(x_2)$.}
\item {\textbf{{3}}}.\ {\sl Denote the support of a variable $X$ as ${\sf supp}[X]$. Let 	
$\mathcal{X}^\circ$ denote the set composed by the interior points of a set $\mathcal{X}$. }
\item {\textbf{{4}}}.\ {\sl Let $\Vert \pi\Vert_p=({\int |\pi(x)|^p d\mu(x)})^{1/p}$ denote the $L_p$-norm, where $\pi$ is a measurable function on a measure space $L(\Omega,\mathcal{B},\mu)$, and $p\in[1,\infty)$.  When $p=\infty$, we denote  $\Vert \pi\Vert_\infty=\inf\ \{c:\mu(\{x:|\pi(x)|>c\})=0\}$.}
\item {\textbf{{5}}}.\ {\sl Suppose that $\{a_n\}_{n=1}^\infty$ and $\{b_n\}_{n=1}^\infty$ are two scalar sequences. Define $a_n\simeq b_n$ when $\lim _{n\rightarrow\infty}a_n/b_n=c$, where $c$ is some nonzero constant.}
\item {\textbf{{6}}}.\ {\sl Suppose that $\{A_n\}_{n=1}^\infty$ and $\{B_n\}_{n=1}^\infty$ are two random variable sequences. Denote $A_n=O_{a.s.}(B_n)$ when there exists a positive constant $M$ satisfying that $|A_n|\leq M\cdot B_n$ almost surely for $n$ large enough, and $A_n=o_{a.s.}(B_n)$ when $\lim _{n\rightarrow\infty}A_n/B_n=0$ almost surely.}
\end{description}
Besides, all asymptotics are discussed when the sample size $n\rightarrow\infty$ without any special statement.

\section{Preliminary and methodology}\label{sec2}
\renewcommand{\theequation}{2.\arabic{equation}}
\setcounter{equation}{0}

We denote the number of change point as $l$. If $l=0$, there is no change. We can write the conventional nonparametric regression model without a change as below,
\begin{equation}\label{m1}
Y_t=\varphi(X_t)+\epsilon_t,1\leq t\leq n,
\end{equation}
where $(X_t,Y_t)$ is the stationary observation at time $t$, $\varphi(x)$ is the conditional mean of $Y_t$ on $X_t=x$, that is,  $\varphi(x)={\sf E}[Y_t|X_t=x]$, and $\epsilon_t$ is the residual, that is, $\epsilon_t=Y_t-{\sf E}[Y_t|X_t]$.

Let $p(x,y)$ and $f(x)$ be the joint density of $(X_t,Y_t)$ and density of $X_t$, respectively. Using the fact that the conditional mean function can be written as
$\varphi(x)=g(x)/f(x)$,
where $g(x)=\int yp(x,y)dy$, one can construct the classical N-W kernel regression estimator (refer to \cite{kr1} and \cite{kr2}) as follows (using the sample from time $s=1$ to $u=n$ in this case),
\begin{equation}\label{eqvarphi}
\widehat \varphi_{s,u}(x;h_n)=\frac{\widehat g_{s,u}(x;h_n)}{\widehat f_{s,u}(x;h_n)},
\end{equation}
with
\begin{equation}\label{eqg}
\widehat g_{s,u}(x;h_n)=\frac{1}{u-s+1}\sum_{t=s}^uY_tK_{h_n}(X_t-x),
\end{equation}
and
\begin{equation}\label{eqf}
\widehat f_{s,u}(x;h_n)=\frac{1}{u-s+1}\sum_{t=s}^uK_{h_n}(X_t-x),
\end{equation}
where $K(\cdot)$ is a kernel function, $h_n$ is a bandwidth, and $K_{h_n}(X_t-x)=1/h_n \cdot K((X_t-x)/h_n)$.
Note that $\widehat g_{s,u}(x;h_n)$ and $\widehat f_{s,u}(x;h_n)$ are the estimators of $g(x)$ and $f(x)$, respectively.

Now we consider the case when there is only one change point, that is, $l=1$.
If the regression function is not invariant in the whole time period but changes after time $k$,
we can create a model as below:
\begin{equation} \label{m2}
Y_t=\left\{
\begin{aligned}
\varphi_1(X_t)+\epsilon_t,\quad 1\leq t \leq k, \\
\varphi_2(X_t)+\epsilon_t,\quad k<t\leq n,\\
\end{aligned}
\right.
\end{equation}
where $\varphi_1\neq\varphi_2$. % $\{X_t,\varepsilon_t\}_{t=1}^n$ is strictly stationary and ${\sf E}[\epsilon_t|X_t]=0$.
Note that at this stage we allow the change of the distribution of $X$ at $k$. In order to identify the model,  an overlapping condition is necessary for the supports of $X$ before and after the change point. However, if the distribution of $X$ changes, the change point detection is much easier, which reduces to a distributional change problem. For this reason, we assume that the distribution of $X_t$ does not change over time. We refer to \cite{np4} for the case that the  distribution of $X_t$ may change as well.
Similarly, we should not only focus on the change of the distribution of $Y$. There are many examples, e.g. $\varphi_1(x)=-\varphi_2(x)=x$, $X_t\sim U[-1,1]$ and $\epsilon_t \sim N(0,1)$, in which it is impossible to detect the regression function change by only detecting the distributional change of $Y_t$.

Next we construct a CUSUM statistic.
Inspired by the statistic due to \cite{np1}, which is defined by
\begin{equation}
\frac{t(n-t)}{n^2}\sup_{x\in \mathbb{R}}\left|\widehat \varphi_{1,t}(x;h_n)-\widehat \varphi_{t+1,n}(x;h_n)\right|,
\end{equation}
we define a CUSUM statistic using sum-of-squares as follows,
\begin{equation}\label{eqw}
W_{1,n}(t)=\frac{t(n-t)}{n^2}\sum_{i=1}^m\left|\widehat \varphi_{1,t}(x_i;h_n)-\widehat \varphi_{t+1,n}(x_i;h_n)\right|^2,
\end{equation}
for $t=1,\cdots,n-1$,
where $\{x_1,\,\cdots,\,x_m\}\subseteq ({\sf supp}[X_1])^\circ$ are $m$ chosen grid points at which the function $\varphi(\cdot)$ is estimated to detect whether a change point exists.

The choice of using supremum or sum-of-squares to aggregate the CUSUM statistic depends on the behavior of the two regression functions. If the magnitude of the change is small in the whole domain of the function, then the cumulation by sum-of-squares is a better choice. If the change of the function is spiky and local, then the  supremum method is better. Thus, the choice of the two methods depends on the model of research and the volatility of data.

We shall also note that when supremum is used to aggregate the CUSUM statistic, the maximum on pre-determined grid points is used in the algorithm for discretization. Therefore, no matter supremum or sum-of-squares is used, we have to introduce the grid points. The requirement for the grid points is that
$\varphi_1(x_i)\neq\varphi_2(x_i)$ for some $1\leq i\leq m$.
In practice, we can use equidistant grid points to cover the whole support of $X_t$. If we have prior information,
 it will be more efficient if we only use the grids points  where the change of the regression function may occur.
For simplicity, we assume that $m$ does not change with $n$. If $m$ increases with $n$, the theoretical results will be more complex and it will be more time-consuming during the computation when $n$ is large.
 We will show in the simulation that the number of $m$ does not matter much to the empirical results.

After we construct the detection statistic, we obtain the estimator for the change point $k$ by maximizing $W_{1,n}(t)$ with respect to $t$, that is,

\begin{equation}\label{eqk}
\widehat {k}= \mathop{\arg\max} \limits_{\Delta_n\leq t \leq n-\Delta_n} W_{1,n}(t),
\end{equation}
where $\Delta_n$ is an integer growing with $n$, which is introduced to deal with the time-boundary effect.
When  $t$ is close to $1$ or $n$, we will have inadequate samples to estimate either
$\varphi_{1,t}$ or $\varphi_{t+1,n}$. Then it will be unable to detect the change by using the CUSUM statistic.
For this reason, we only detect change points which is not close to the time boundary.
We note that the weight $t(n-t)/n^2$ in the CUSUM statistic, similar to \cite{np1}, is also introduced to deal with the time-boundary effect.
These two strategies are commonly used, see discussion in Appendix D of \cite{BCF18} for $\Delta_n$
and \cite{th2} for an alternative choice of weights.

\section{Asymptotics}\label{sec3}
\renewcommand{\theequation}{3.\arabic{equation}}
\setcounter{equation}{0}
\newtheorem{assumption}{Assumption}
Before introducing the related theorems, we will  make the following assumptions for the purpose of asymptotic analysis.

\begin{assumption}\label{a1}
(PSM) The  process $\{X_t,\epsilon_t\}_{t=1}^n$ is strictly stationary and $\{X_t,Y_t \}_{t=1}^n$ is strongly mixing dependent with the mixing coefficient $\alpha(t)$ decaying to zero at a polynomial rate, that is, $\alpha(t)\leq C_1|t|^{-\gamma}$ with some constant  $\gamma> 11/2$ and constant $C_1>0$.
\end{assumption}

\begin{assumption}\label{a2}
(GSM) The process $\{X_t,\epsilon_t\}_{t=1}^n$ is strictly stationary and $\{X_t,Y_t \}_{t=1}^n$ is strongly mixing dependent with the mixing coefficient $\alpha(t)$ decaying to zero at a geometric rate, that is, $\alpha(t)\leq C_1\rho^{t}$ with some constant $0<\rho< 1$ and constant $C_1>0$.
\end{assumption}

\begin{assumption}\label{a3}
Suppose that $h_n\simeq n^{-\omega}$, where $\omega$ satisfies that
$0<\omega<1-\frac{14}{2\gamma+3}$ for PSM, or $0<\omega<1$ for GSM.
\end{assumption}

\begin{assumption}\label{a4}
Let the kernel function $K$ be a symmetric nonnegative function satisfying that $0< \|K\|_1<\infty $ and $0<\|K\|_\infty<\infty$.
\end{assumption}

\begin{assumption}\label{a5}
Suppose that $f$ and $g$ defined in the model (\ref{m1}) are uniformly bounded and denote $f^{(t)}$ as the joint density of $(X_t,X_0)$. Define the bivariate function $F^{(t)}=f^{(t)}-f\bigotimes f$, and
   $${\sf G}^{(t)}(x_t,x_0)=\int|y_ty_0p(y_t,y_0,x_t,x_0)-y_tp(y_t,x_t)y_0p(y_0,x_0)|dy_tdy_0,$$
    where $p(y_t,y_0,x_t,x_0)$ and $p(y_t,x_t)$ are the joint densities of $ (Y_t,Y_0,X_t,X_0)$ and $(Y_t,X_t)$, respectively.
   Assume that for some constants $p_F>2$ and $p_G>2$, $\sup_{t\in\mathbb{Z}_+}{\|F^{(t)}\|}_{p_F}<C_2<\infty$ and $\sup_{t\in\mathbb{Z}_+}{\|G^{(t)}\|}_{p_G}<C_3<\infty$ . In addition, if the data is PSM, we define $q_F:=1-2/p_F$ and $q_G:=1-2/p_G$, and require that $q_F(\gamma-1)>1$ and $q_G(\gamma-1)>1$, where $\gamma$ is defined in Assumption \ref{a1}.
\end{assumption}

\begin{assumption}\label{a6}
 Suppose that $\sup \limits_{x\in {\sf supp}[X_t]}{\sf E}\left[Y_t^{2}|X_t=x\right]<C_4$ and  ${\sf E}\left[e^{C_5|Y_t|}\right]<C_6$, where $C_4, C_5$ and $C_6$ are positive constants.
\end{assumption}

If there is a change point, $p(y,x)$ and $g(x)$ may change over time. Assumptions 5 and 6 should hold true separately on the subsamples before and after the change point.
In addition, if a change happens, we assume that:
\begin{assumption}\label{a7}
Let the time-scaled change point located at $k=\lfloor\theta n\rfloor$ with $\delta\leq\theta\leq1-\delta$ for some $\delta>0$.
Let $\mathcal{X}= ({\sf supp} [X_1])^\circ$ and $\mathcal{Y}=\{x: \varphi_1(x)\neq\varphi_2(x)\}^\circ$. We assume that $\mathcal{X}\cap \mathcal{Y}$ is non-empty, and $\varphi_1(x)$ and $\varphi_2(x)$, which are two bounded functions, do not change with $n$.
\end{assumption}

\begin{assumption}\label{a8}
The grid points $\{x_i\}_{i=1}^m$ satisfy that $\{x_i\}_{i=1}^m\subset \mathcal{X}$ and $x_i\in \mathcal{Y}$ at least for some grid point $x_i$. The number of grid points $m$ is not dependent on $n$.
\end{assumption}
\smallskip

The condition of the mixing dependence on $\{X_t,Y_t\}_{t=1}^n$ in Assumption \ref{a1} (or \ref{a2}) is very common and covers some commonly-used time series models such as ARMA process (\cite{Bo98}) and GARCH process (\cite{garch}).
\cite{Bo98} investigated the nonparametric N-W estimation for GSM and \cite{JR11} extended it to PSM. The condition that $\{X_t,\epsilon_t\}_{t=1}^n$ is stationary guarantees that $\{X_t,Y_t\}_{t=1}^n$ is stationary before and after the change, respectively.
Assumptions \ref{a3} is some restrictions on the bandwidth.
Given the order of $h_n$, it holds true that $\log^2n/(nh_n)\rightarrow 0$, which has been used in Chapter 2 of \cite{Bo98} to obtain the upper bound of $v^2(q)$ defined in Lemma \ref{ppst:mixing}.
The condition on the kernel function in Assumption \ref{a4} is very mild, which is satisfied by commonly-used kernels.
Assumption \ref{a5} is a technique requirement which refers to the same assumption in \cite{Bo98} and \cite{JR11}, and it is  used by us to bound the variance and covariance of $K_{h_n}(X_t-x)$ and $Y_tK_{h_n}(X_t-x)$ together with Assumption \ref{a6}.
If there is no change point,
Assumptions \ref{a3}--\ref{a6} guarantee the uniform bound of $\widehat f-{\sf E}[\widehat f]$ and $\widehat g-{\sf E}[\widehat g]$. %We use it to obtain the order of $W_{1,n}(t)$.

In Assumption \ref{a7},
%the condition  $\inf_{x\in\mathcal{X}}f(x)>C_7>0$,
%which has been used in many papers such as the Condition 8 of \cite{densityinf}, can be applied to many distributions such as normal, uniform, gamma, and beta distribution.
what one should note is that we do not have to require that $f$, $g$ or $\varphi$ are continuous or differentiable as many researchers did.
If $\varphi_1(x)$ and $\varphi_2(x)$  are discontinuous functions,  the local estimator estimates a smoothing version of $\varphi_1(x)$ and $\varphi_2(x)$. We can still detect the change point by comparing the two estimates. %Generally, we even do not require an consistent estimate of $\varphi(x)$.
Assumption 8 suggests that, if a change happens, it is enough for our detection method to locate the change point, as long as there is a grid point on which the values of $\varphi_1$ and $\varphi_2$ are not identical. This quality releases the selection of the grid points extremely. As we have mentioned before, we can choose the equidistant grid points covering the main region between the minimum and the maximum of the collected $\{X_t\}_{t=1}^n$. In this case, the grid points can cover each local region and probably some of them lie in $\cal Y$.
%Instead, we adopt the condition on $\Omega_n$, which is used to obtain the nonzero lower bound of  $\Lambda_n$ defined in Theorem \ref{theorem2}, and derive the nonzero lower bound of $\max_t W_{1,n}(t)$ (see Theorem \ref{theorem2}) further when a change happens. Indeed, unlike traditional kernel smooth methods, we allow the discontinuity in the regression functions before and after the change point. After the change point is located,  the discontinuity problem can be solved easily on each sub-sample using existed methods. Thus, the discontinuity detection is not what we are going to investigate.  Note that the idea that ${\sf E}[\widehat \varphi]-{\sf E}[\widehat g]/{\sf E}[\widehat f]=o(1) $ from Theorem 1 of \cite{kr1}, although under different regularity conditions with ours, heuristically, that $\Lambda_n$ is bounded away from zero can be seen as a condition of nonzero lower bound on $\sum_{i=1}^m ({\sf E}[\widehat\varphi_{1,t}(x_i)-\widehat\varphi_{t+1,n}(x_i)])^2$. In other word, we require that the structural change is not too tiny ($\Lambda_n$ has nonzero lower bound).
\bigskip

Recall that, if $l=0$, it means that no change appears, and the structure is of form as the model (\ref{m1}). If $l=1$, it means that a change appears, and the structure is of form as the model (\ref{m2}). We have the following results for these two cases.
\begin{theorem}\label{theorem1}
Suppose that the process $\{X_t,Y_t \}_{t=1}^n$ is PSM (or GSM) and Assumptions \ref{a3}--\ref{a6} are satisfied. Let $ W_{1,n}(t)$ be defined in (\ref{eqw}). Then we have when $l=0$,
\begin{equation}
\max_{\Delta_n\leq t\leq n-\Delta_n}W_{1,n}(t)=O_{a.s.}\left(\frac{\log ^4 n}{nh_n}\right),
\end{equation}
where $\Delta_n=\lfloor n\delta\rfloor$ for some constant $\delta$.
\end{theorem}
\smallskip

When there is a change point, define $$\Lambda_{h_n}(x):=\frac{{\sf E}[Y_{k}K_{h_n}(X_{k}-x)]}{{\sf E}[K_{h_n}(X_{k}-x)]}
-\frac{{\sf E}[Y_{k+1}K_{h_n}(X_{k+1}-x)]}{{\sf E}[K_{h_n}(X_{k+1}-x)]}
,$$
for $x\in {\cal X}$.
Indeed, if $f$ and $g$ are differentiable functions, we have by integration and Taylor expansion that$$\lim_{h_n\to 0}\Lambda_{h_n}(x)=\lim_{h_n\to 0}\frac{{\sf E}[\varphi_1(X_k)K_{h_n}(X_{k}-x)]}{{\sf E}[K_{h_n}(X_{k}-x)]}
-\lim_{h_n\to 0}\frac{{\sf E}[\varphi_1(X_{k+1})K_{h_n}(X_{k+1}-x)]}{{\sf E}[K_{h_n}(X_{k+1}-x)]}=\varphi_2(x)-\varphi_1(x).$$
Thus $\Lambda_{h_n}(x)$ can be seen as a smooth version of $\varphi_2(x)-\varphi_1(x)$.
We have the following theorem.

\begin{theorem}\label{theorem2}
 Suppose that the process $\{X_t,Y_t \}_{t=1}^n$ is PSM (or GSM) and Assumptions \ref{a3}--\ref{a8} are satisfied. Let $ W_{1,n}(t)$ be defined in (\ref{eqw}). Then when $l=1$ and $n$ is large enough, we have $\Lambda_{h_n}(x_i)$ is bounded away from zero if   $x_i\in \mathcal{X}\cap\mathcal{Y}$.
 What's more,
\begin{eqnarray}\label{eqT2}
\max\limits_{\Delta_n\leq t\leq n-\Delta_n}W_{1,n}(t)\geq  \theta(1-\theta)\sum_{i=1}^m\Lambda_{h_n}^2(x_i)+O_{a.s.}\left(\frac{\log ^2 n}{\sqrt{nh_n}}\right)
\end{eqnarray}
where $\Delta_n=\lfloor n\delta\rfloor$ and $\theta$  is defined in Assumption \ref{a7}.
\end{theorem}
We can see from Theorem \ref{theorem2} that the CUSUM statistic is larger if $\sum_{i=1}^m\Lambda_{h_n}^2(x_i)$ is larger or if $\theta$ is close to 0.5. This means that it is easier to detect a change point if the change is larger in magnitude, or if the change point locates in the middle of the sample set.

\bigskip

With probability one, Theorem \ref{theorem1} shows the asymptotic order of the CUSUM statistic when  $l=0$. There exists a constant $M$ satisfying that $\max_t W(t)\leq M\log ^4 n/(nh_n)$ asymptotically. Theorem \ref{theorem2} shows that our CUSUM statistic has an asymptotic lower bound when $l=1$, and this lower bound is a constant and thus is larger than the order in Theorem \ref{theorem1} as $\log ^4 n/(nh_n)\rightarrow 0$.
Therefore, we can find an appropriate threshold $\xi_n$ between $M\log^4n/(nh_n)$ and $\theta(1-\theta)\sum_{i=1}^m\Lambda_{h_n}^2(x_i)$ when $n$ is large enough.
Since both $\theta$ and $\Lambda_{h_n}^2(x_i)$ are unobservable, we will use a permutation method to set the threshold for our method in the simulation and real data analysis.

We determine the value of $l$ by comparing  $\max_t W(t)$ with the threshold $\xi_n$.
When $\max_t W(t)\leq \xi_n$, we set $\widehat l=0$, otherwise $\widehat l=1$.
Then we derive the strong consistency of $\widehat \theta:=\widehat k/n$ as follows.

\begin{theorem}\label{theorem3}
Suppose that the conditions in either Theorem \ref{theorem1} or Theorem \ref{theorem2} are satisfied,  if the threshold is between $M\log^4n/(nh_n)$ and $\theta(1-\theta)\sum_{i=1}^m\Lambda_{h_n}^2(x_i)$,  we have that\\
(i) $\widehat l=l$
almost surely when $n\rightarrow\infty$. \\
(ii) When $l=1$,
$|\widehat \theta -\theta|=o_{a.s.}(1)$,
where $\widehat \theta =\widehat k/n$, $\widehat k$ is defined in (\ref{eqk}), and $\theta$ is defined in Assumption \ref{a7}.
\end{theorem}
\smallskip

We will show the determination of the threshold in the simulation.

%So far, we have not introduced smooth conditions on the regression functions. Actually, if we suppose that $f$ and $g$ are differentiable as many researchers do, we can eliminate the condition on $\Omega_n$ and also obtain the strong consistency as Corollary \ref{corollary1} shows. At this moment, we only require that $\varphi_1$ is not equal to $\varphi_2$ at some grid point. That is, it is enough for our detection as long as the two functions are different at one grid point. The detectable change can be smaller than before.
%\smallskip

%\begin{corollary}\label{corollary1}
% Suppose that the process $\{X_t,Y_t \}_{t=1}^n$ is PSM (or GSM), Assumptions \ref{a3}--\ref{a6} are satisfied, and $f(x), g(x)$ and $p(x,y)$ are differentiable. Let change point be $k$ and $ W_{1,n}(t)$ be defined as (\ref{eqw}). Then we have (i) $\widehat l=l$
%almost surely when $n\rightarrow\infty$. (ii) When $l=1$
%\begin{equation}
%	|\widehat \theta -\theta|=o_{a.s.}(1),
%\end{equation}
%if $\inf\limits_{x\in\mathcal{X}}f(x)>C_7>0$, and $\exists$ $x_i \in \mathcal{X}^\circ $ such that $\varphi_1(x_i)\neq\varphi_2(x_i)$
%, where $\widehat \theta =\widehat k/n$, $\widehat k$ is defined in (\ref{eqk}), and $\mathcal{X}$ and $\theta$ are defined in Assumption \ref{a7}.
%\end{corollary}

\section{Practicalities and simulation studies}\label{sec4}
\renewcommand{\theequation}{4.\arabic{equation}}
\setcounter{equation}{0}
\renewcommand{\thetable}{\textbf{\arabic{table}}}
\setcounter{table}{0}
\renewcommand {\thefigure} {\textbf{\arabic{figure}}}
\setcounter{figure}{0}

In Section \ref{sec4.1}, we discuss the choice of the bandwidth $h_n$ and the threshold $\xi_n$. We compare the estimation precision of several detection methods in Section \ref{sec4.2}. The Epanechnikov  kernel function ${K}(u)=3/4(1-u^2)I(|u|\leq 1)$ is used in the N-W estimation throughout the whole simulation.
In terms of the grid points, which are related to  $\mathcal{X}$ and $\mathcal{Y}$ in Assumption \ref{a7}, we choose $20,50,100$ equidistant points between 5th percentile and 95th percentile of $\{X_t\}_{t=1}^{n}$ as the grid points, respectively, to construct the CUSUM statistic in (\ref{eqw}). We figure out that the results are similar to each other and thus set $m=100$ hereafter.

% Note that N-W estimator tends to  have big estimation bias around the support boundary (see \cite{npref}), which may result in a large value of $W_{1,n}(t)$ even if there is no change point. we trim the two intervals around the boundary and use the interval from 5th percentile to 95th percentile of $\{X_t\}_{t=1}^n$.

Two data generation processes (DGP), the ARMA process and the ARFIMA process, are used throughout our simulation.
The former is a strongly mixing process, see the argument in S.4 of \cite{Bo98}, which matches Assumption \ref{a2} in  Section \ref{sec3}.
The latter is not a strongly mixing process but a long-memory one which \cite{np1} has studied. That helps us to show the robust application of our proposed method.
The long-memory assumption together with ARFIMA DGP  (see \cite{arfima81} for detail) is introduced to the nonparametric change-point-detection problem by \cite{np1}.
Because \cite{np1}'s method is based on the N-W estimation as we do but uses supremum to construct the CUSUM statistic, we denote it as ``nwsup''. In contrast, our method is denoted as ``nwss'', where ``ss'' means ``sum-of-squares''.

During our research, we ever planed to replace the N-W estimator with the local linear estimator in constructing the CUSUM statistic.
%, that is,
%$$
%W_{1,n}(t)=\frac{t(n-t)}{n^2}\sum_{i=1}^m\left|\widehat \varphi_{1,t}(x_i;h_n)-\widehat \varphi_{t+1,n}(x_i;h_n)\right|^2,
%$$
%where $\widehat\varphi_{1,t}$ and $\widehat\varphi_{t+1,n}$ are estimated by the local linear estimation using the sample sets $\{X_1,\cdots,X_t\}$ and $\{X_{t+1},\cdots,X_n\}$, respectively.
 We replace the N-W estimators in ``nwsup'' or ``nwss'' by the local linear estimators in the change-point-detection procedure to obtain the new methods which we denote as ``llsup'' and ``llss'', respectively.
%We will compare these methods in our simulation.
Nevertheless, we will see later in our simulation that it does not promote performance result, compared with our current method.

Two nonparametric structural changes are considered:

(i) The first DGP is
\begin{eqnarray}\label{model1}
Y_t=\left\{
\begin{aligned}
1+X_t+\epsilon_t,1\leq t\leq k,\\
X_t^2+\epsilon_t\ \ \ ,k<t\leq n.\\
\end{aligned}
\right.
\end{eqnarray}
The relation between $X_t$ and $Y_t$ changes from a linear pattern to a quadratic one.
 This DGP is constructed to show the detectable property of our method even if ${\sf E}[Y_t]$ does not change.

(ii) To show the extent of detectable structural change, the second DGP allows different scales of the change of  $\varphi(x)$, which is given by
\begin{eqnarray}\label{model2}
 Y_t=\left\{
\begin{aligned}
X_t^2+\epsilon_t\ \ \ \ \ \ ,1\leq t\leq k,\\
(X_t+\Delta_{\varphi})^2+\epsilon_t,k<t\leq n,\\
\end{aligned}
\right.
\end{eqnarray}
where $\Delta_{\varphi}$ determines the magnitude of the change of the regression function.

For each model, we simulate $X_t$ by an ARMA(1,1) process or an ARFIMA(0,0.15,0) process:

(i) The ARMA(1,1) process is given by
\begin{equation}
(1-0.5{\cal L})X_t=(1+0.5{\cal L})u_t,
\end{equation}
where ${\cal L}$ is the time-shifting operator, $u_t$ is generated from an
independent normal distribution with mean 0 and variance $3/7$,
and $\epsilon_t$ is  generated from an
independent normal distribution with mean 0 and variance 0.25.
Note that in this case ${\sf E}[X_t]=0$ and ${\sf E}[X_t^2]=1$, therefore ${\sf E}[Y_t]$ in the model (\ref{model1}) does not change over time $t$ but changes in model (\ref{model2}).

(ii) The ARFIMA(0,0.15,0) process is given by
\begin{equation}\label{eq.4.4}
(1-{\cal L})^{0.15}X_t=u_{1t},
\end{equation}
where $u_{1t}$ is generated from an
independent normal distribution with mean 0 and variance $\Gamma^2(0.85)/\Gamma(0.7)\approx 0.9534$. In the meantime, $\epsilon_t$ is  generated from an
ARFIMA(0,0.35,0) process as follows,
\begin{equation}\label{4.5}
(1-{\cal L})^{0.35}\epsilon_t=u_{2t},
\end{equation}
where $u_{2t}$ is generated from an
independent normal distribution with mean 0 and variance 0.01.
The setting of variance still leads that ${\sf E}[X_t]=0$ and ${\sf E}[X_t^2]=1$ (one can refer to the statement about the variance of ARFIMA process in \cite{arfima81}). Thus  ${\sf E}[Y_t]$ in the model (\ref{model1}) does not change over time $t$, while changes slightly in the model (\ref{model2}).

To quantify the performance of each method, before exhibiting our simulation, we introduce some notations. Let the change point estimate of the $i$-th experiment be $\widehat k_i$, for $i=1,\cdots,N$, where $N$ is the number of experiments. We show the consistency of the estimator by using ``Bias'', which is defined as $N^{-1}\sum_{i=1}^N(\widehat k_i-k)$. We compare different methods by using absolute bias (ABias), which is defined as $N^{-1}\sum_{i=1}^N|\widehat k_i-k|$.
The sample standard deviations of  $\widehat k_i-k$ and $|\widehat k_i-k|$ are denoted as ``BiasSd'' and ``ABiasSd'', respectively.

\subsection{Selection of bandwidth and threshold} \label{sec4.1}
The bandwidth used in the change point detection can be different from that used in the nonparametric regression. Our aim is not to find a better regression curve, say in the sense of least MSE or some other criteria, but to make the change most detectable.
In the case when the unconditional mean of $Y_t$ changes, one can always benefit from choosing a large $h_n$. This is because the nonparametric curve degenerates to a horizontal line when $h_n$ tends to infinity, and detecting the change point in the conditional mean is actually a parametric problem which naturally has a better solution in the sense of the asymptotics.

\begin{figure}[H]
\centering
\subfigure[F(h) using ARMA DGP under model (4.1)]{\includegraphics[width=0.55\textwidth]{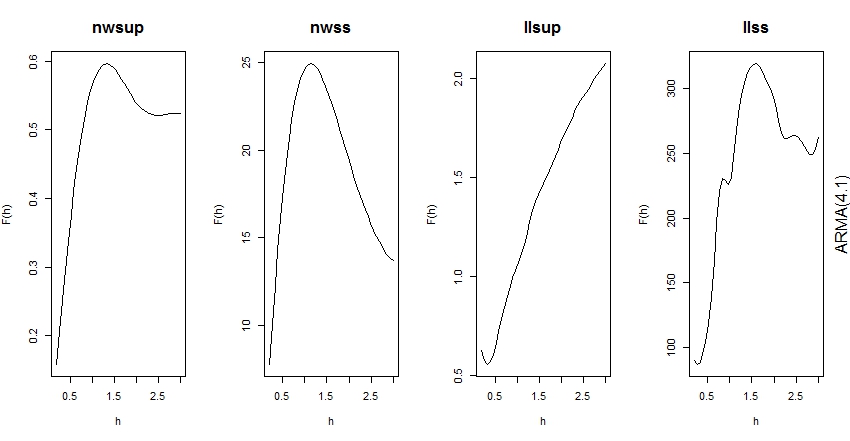}}
\subfigure[ABias using ARMA DGP under model (4.1)]{\includegraphics[width=0.55\textwidth]{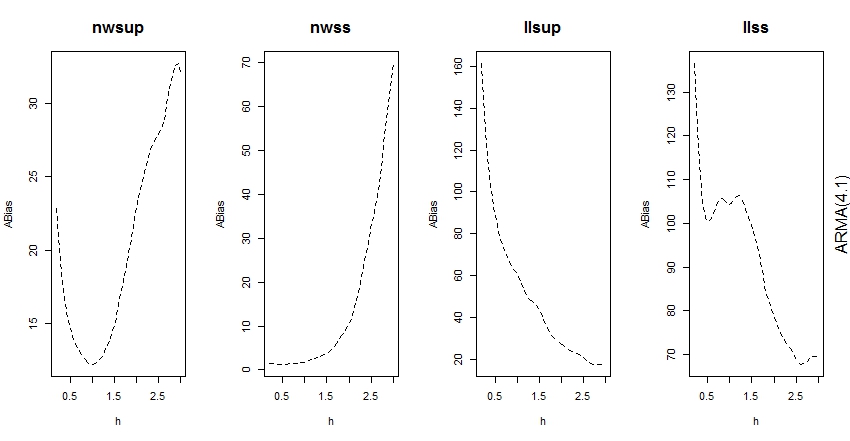}}
\subfigure[F(h) using ARMA DGP under model (4.2)]{\includegraphics[width=0.55\textwidth]{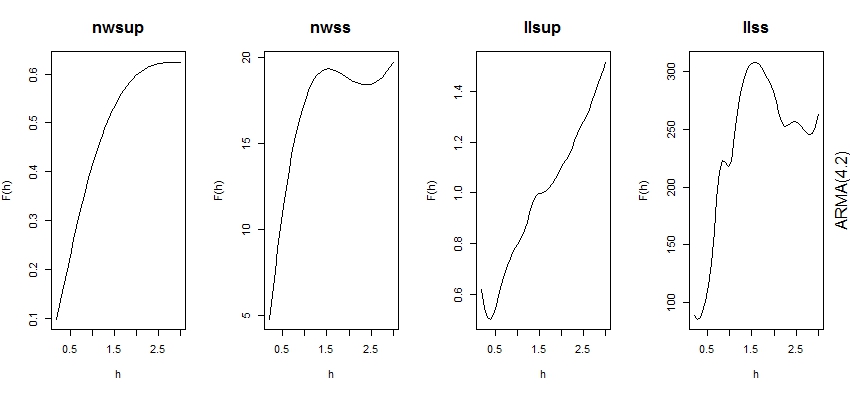}}
\subfigure[ABias using ARMA DGP under model (4.2)]{\includegraphics[width=0.55\textwidth]{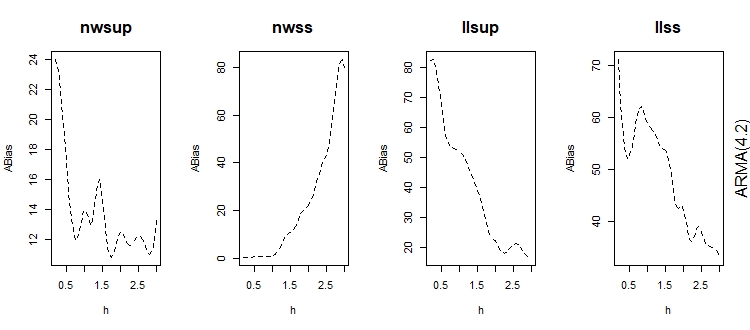}}
\caption{\ Simulating $X_t$ by ARMA(1,1) process and $\epsilon_t$ by N(0,0.25), we obtain the F(h) and ABias of each method. Note that each column corresponds to a detection method, and the first two rows demonstrate the experimental F(h) and ABias under the model (\ref{model1}) while the last two rows show those under the model (\ref{model2}) with $\Delta_{\varphi}=0.5$ without loss of generality.}\label{ARMAFhABias}
\end{figure}

\begin{figure}[H]
\centering
\subfigure[F(h) using ARFIMA DGP under model (4.1)]{\includegraphics[width=0.55\textwidth]{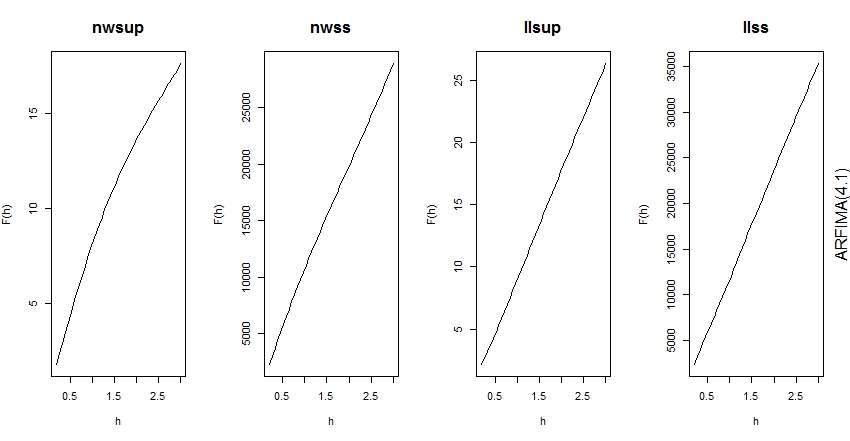}}
\subfigure[ABias using ARFIMA DGP under model (4.1)]{\includegraphics[width=0.55\textwidth]{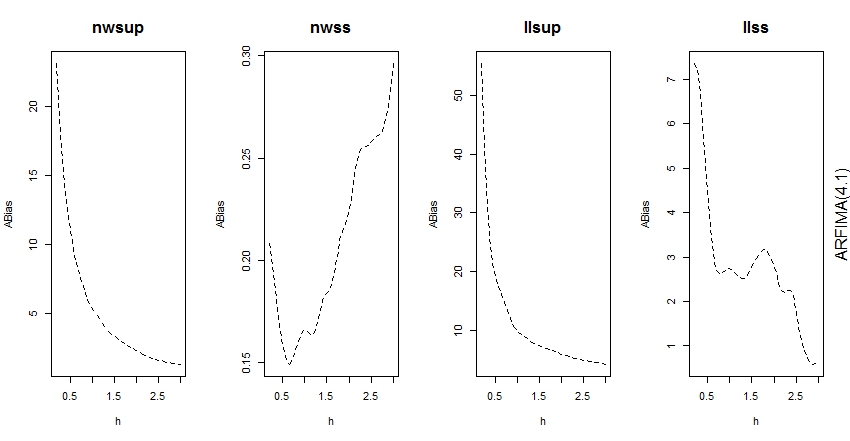}}
\subfigure[F(h) using ARFIMA DGP under model (4.2)]{\includegraphics[width=0.55\textwidth]{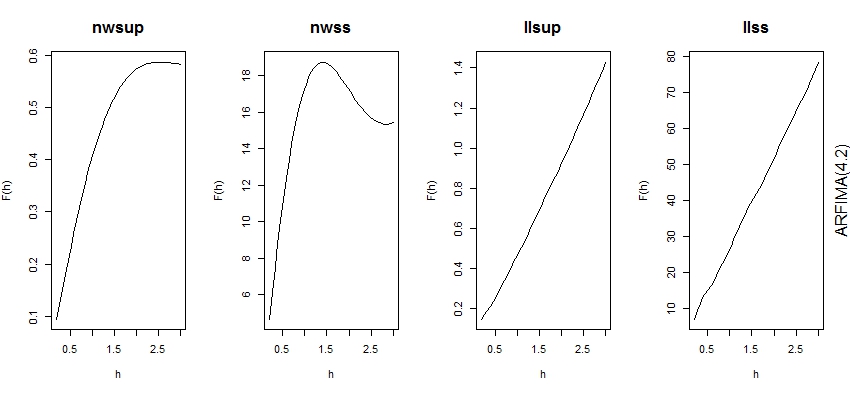}}
\subfigure[ABias using ARFIMA DGP under model (4.2)]{\includegraphics[width=0.55\textwidth]{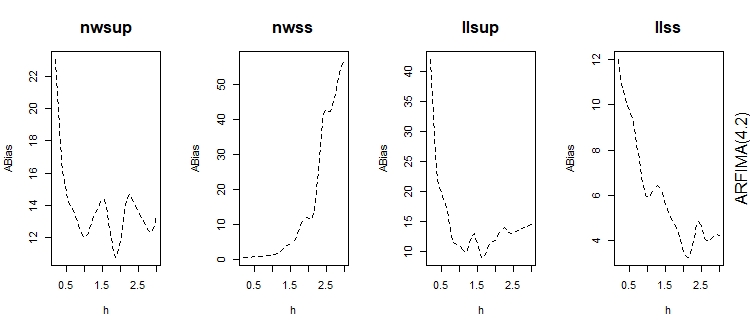}}
\caption{\ Simulating $X_t$ by ARFIMA(0,0.15,0) process and $\epsilon_t$ by ARFIMA(0,0.35,0) process, we obtain the F(h) and ABias of each method. Note that each column corresponds to a detection method, and the first two rows demonstrate the experimental F(h) and ABias under the model (\ref{model1}) while the last two rows show those under the model (\ref{model2}) with $\Delta_{\varphi}=0.5$ without loss of generality.}\label{ARFIMAFhABias}
\end{figure}

We want to find a data-driven procedure to select a bandwidth before knowing whether the  unconditional mean of $Y_t$ changes.
We may find some clues from Theorem \ref{theorem1} and \ref{theorem2}. When there is no change point, the order of $\max_t W_{1,n}(t)$ is $\log^4n/(nh)$. %When there is a change point, however, the specific relation between $\max_t W_{1,n}(t)$ and $h_n$ is not clear because of $C_7$.
The first idea one could think of is to find the $h$ that maximize $\max_t W_{1,n}(t)$. However, when there is no change point, this criterion leads to a tiny bandwidth. %which can be seen from Theorem \ref{theorem1}.
To fix it, we maximize $F(h)=\max_t W_{1,n}(t)\cdot h$ to find a bandwidth.
We have $F(h)\leq M\cdot \log^4n/n$ a.s. from Theorem \ref{theorem1}, where $M$ is a positive constant, showing that the order of $F(h)$ is no longer dependent on $h$. Indeed, the procedure can be seen as trying to find the potential $M$ for the inequality. We avoid using too large $h$ and only consider  $h$ smaller than
$(\max_{1\leq t\leq n }(X_t)-\min_{1\leq t\leq n }(X_t))/2$.
In Figure \ref{ARMAFhABias} and \ref{ARFIMAFhABias}, we show that how the maximum point of $F(h)$ and  the minimum point of ``ABias'' match each other under different models and data.
Specifically, we set sample size $n=500$, $\Delta_{\varphi}=0.5$ in the model (\ref{model2}) and the relative position $\theta=0.4$,  and replicate experiments $N=500$ times for each $h$. The average $\max_t W_{1,n}(t)\cdot h$ of $500$ experimental results is considered as $F(h)$ for each $h$. Note that we do not consider the threshold  temporarily and determine the maximizer of $W_{1,n}(t)$ as the change point.

In most cases, the maximizer of $F(h)$ is near the minimizer of ABias. Particularly, when $F(h)$ increases monotonically, ABias tends to decrease monotonically or retains in a range of low ABias. That is, we can take the
maximizer of $F(h)$ as the bandwidth to ensure the least or low ABias, which we cannot compute in reality. It seems to justify our bandwidth-choosing strategy. Thus, we use this method to determine the bandwidth of the real data demonstrated in the next section. In our simulation, without loss of generality, we take $h=1$ to proceed our simulation since each method has a relatively low ABias.

Now, we use bootstrap to determine the threshold $\xi_n$ of our method ``nwss''. Specifically, we resample 200 sets of samples without replacing (which is a permutation essentially) and compute the 200 values of $\max_t W_{1,n}(t)$. Then we define the 99th percentile of $\max_t W_{1,n}(t)$'s as the threshold, which is an approximation of the order in Theorem \ref{theorem1}. To show the performance of our detection and threshold determination, we replicate 500 detection experiments
for each specification defined in Section \ref{sec4.1}.
For each case, the percentage of experiments determining a change happens, denoted as PDC, is shown in Table \ref{PDC}, that is, the percentage of $\max_t W_{1,n}(t)>\xi_n$. PDC shows the rate of the true positive results when there is a change point or the rate of the false positive results when the system is stable and it can measure the detection performance well.

From the table, we can see that: (i)  When the structural change is large, e.g., the model (\ref{model1}) and the model (\ref{model2}) with $\Delta_{\varphi}=0.3$ or 0.5, our method can detect nearly all the changes.
(ii) When there is no change point, the rate of the false positive results is approximately 2\% for strongly mixing process.
(iii) The accuracy increases with the sample size.
(iv)  When the change point is not around the center ($\theta=0.2$), the rate of the true positive results is slightly inferior than that when $\theta=0.4$ but remains good as long as the sample size is relatively large. Note that the results are the same for $\theta=0.4$ and $\theta=0.2$ when $\Delta_{\varphi}=0$.

These results are in compliance with the theoretical results.
The larger the sample size is, the more accuracy our method is.
It is possible to repeat the resample in the bootstrap procedures more times and select a larger threshold, for example 99.9th percentile.
When sample size is large enough, we are able to find a proper threshold. According to Theorem \ref{theorem1} and \ref{theorem2}, $\max_{t}W_{1,n}(t)$ should be less than the threshold  if there is no change and exceed the threshold when the change point exists.

We conclude that, although our research detection is performed by threshold method rather than hypothesis testing, our method still has a nice rate of the true positive  and a low rate of the false positive  to distinguish whether there is a change point.
\begin{table}[H]
\centering
\scalebox{1}{
\begin{tabular}{c|c|ccc|ccc}
\toprule
\multicolumn{1}{c|}{\multirow{2}*{$X_t$}}&\multicolumn{1}{c|}{\multirow{2}*{model}}& \multicolumn{3}{c|}{PDC($\theta=0.4$)}&\multicolumn{3}{c}{PDC($\theta=0.2$)}\\
\cline{3-8}
\multicolumn{1}{c|}{}&\multicolumn{1}{c|}{}&  $n=200$& $n=500$& $n=1000$&$n=200$& $n=500$& $n=1000$\\
\cline{1-8}
\multicolumn{1}{c|}{\multirow{5}*{ARMA}}&4.1& 0.982&1&1&0.940&1&1\\
\multicolumn{1}{c|}{}&$4.2(\Delta_{\varphi}=0.5)$& 0.888&0.998&1&0.722&0.956&1\\
\multicolumn{1}{c|}{}&$4.2(\Delta_{\varphi}=0.3)$& 0.678&0.924&1&0.562&0.748&0.942\\
\multicolumn{1}{c|}{}&$4.2(\Delta_{\varphi}=0.1)$& 0.036&0.264&0.108&0.014&0.234&0.064\\
\cline{2-8}
\multicolumn{1}{c|}{}&$4.2(\Delta_{\varphi}=0)$& 0.018&0.012&0.016&0.018&0.012&0.016\\
\cline{1-8}
\multicolumn{1}{c|}{\multirow{5}*{ARFIMA}}&4.1&1&1&1&0.992&1&1\\
\multicolumn{1}{c|}{}&$4.2(\Delta_{\varphi}=0.5)$&0.822&1&1&0.426&1&1\\
\multicolumn{1}{c|}{}&$4.2(\Delta_{\varphi}=0.3)$&0.132&0.986&1&0.090&0.700&0.998\\
\multicolumn{1}{c|}{}&$4.2(\Delta_{\varphi}=0.1)$&0.056&0.102&0.082&0.046&0.060&0.052\\
\cline{2-8}
\multicolumn{1}{c|}{}&$4.2(\Delta_{\varphi}=0)$&0.042&0.030&0.024&0.042&0.030&0.024\\
\bottomrule
\end{tabular}}
\caption{\ PDC under different models and data.}\label{PDC}
\end{table}
\bigskip
\subsection{ARMA and ARFIMA processes with a change point}\label{sec4.2}

After judging whether there is a change point, to compare the estimation precision of the four methods,``nwss'', ``nwsup'', ``llss'' and ``llsup'', we do additional simulations under different model settings as below.

\begin{figure}
\centering
\subfigure[$\theta=0.4,n=500$, and ARMA DGP under model (4.1)]{\includegraphics[width=0.35\textwidth]{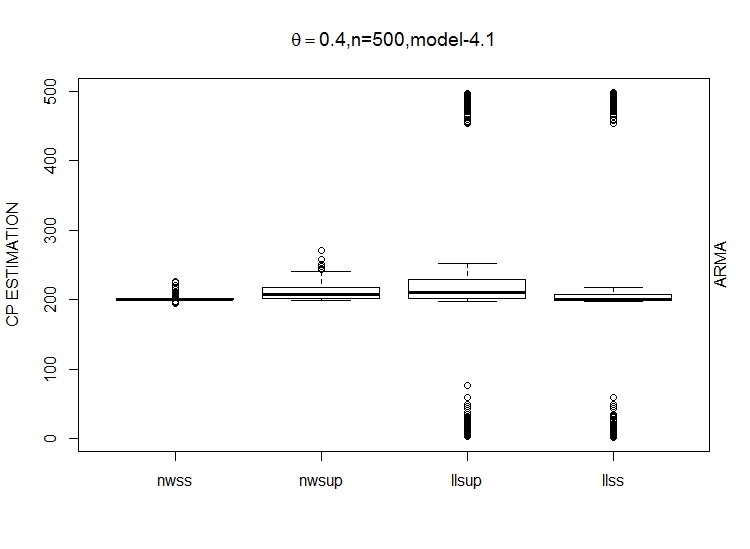}}\\
\subfigure[$\theta=0.4,n=500$, and ARMA DGP under model (4.2)]{\includegraphics[width=0.35\textwidth]{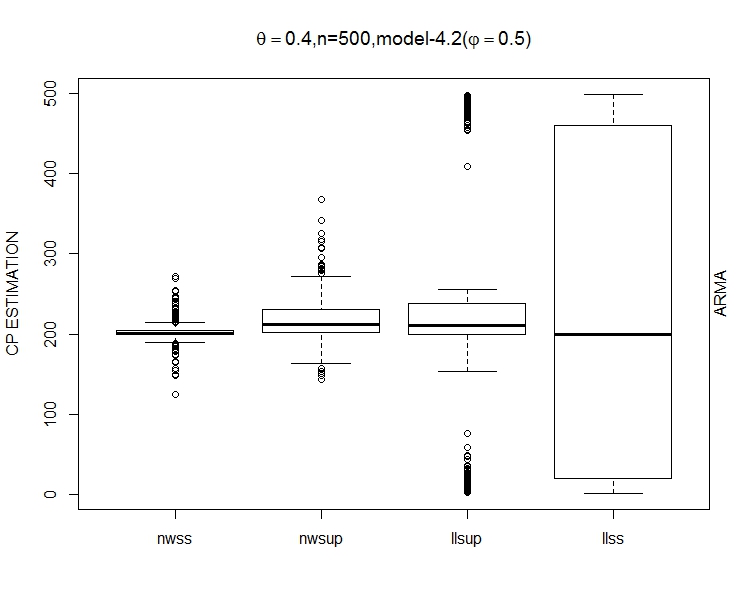}}\\
\subfigure[$\theta=0.2,n=500$, and ARMA DGP under model (4.1)]{\includegraphics[width=0.35\textwidth]{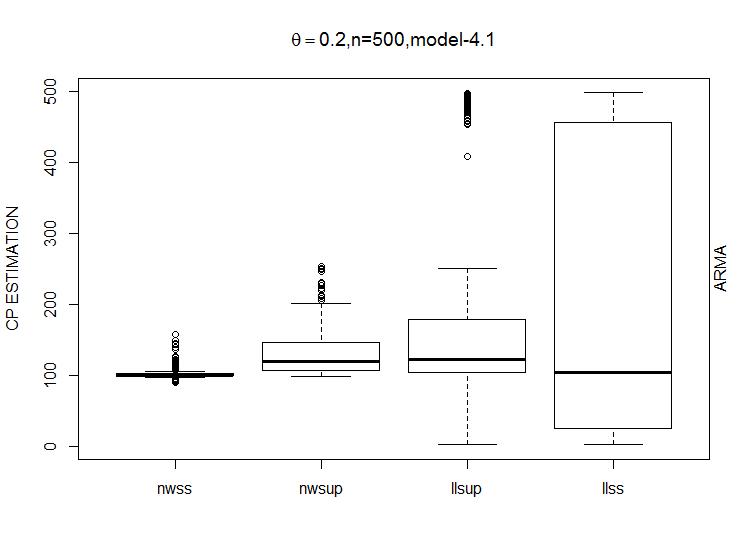}}\\
\subfigure[$\theta=0.2,n=500$, and ARMA DGP under model (4.2)]{\includegraphics[width=0.35\textwidth]{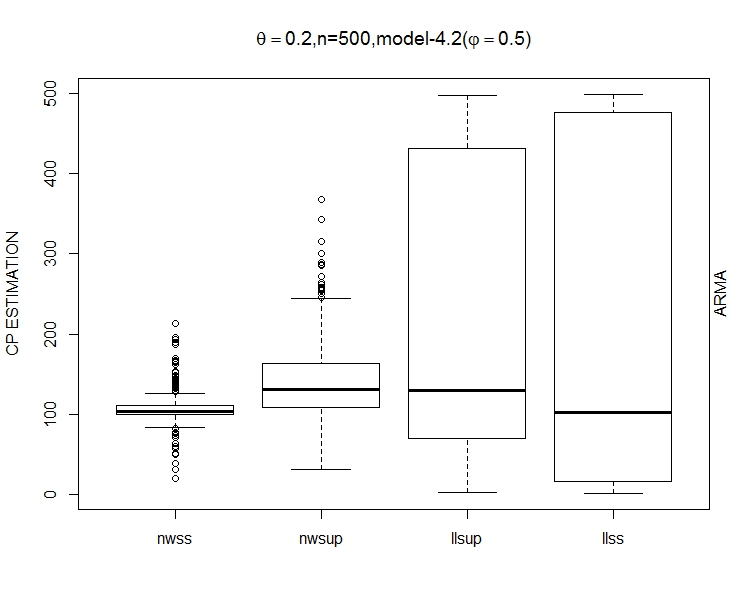}}
\caption{\ At the case when data is generated by ARMA DGP, setting sample size $n=500$, we plot the box-plots of the change point estimates under the model (\ref{model1}) and the model (\ref{model2}) with $\theta=0.4$ and $\theta=0.2$, respectively.}\label{ARMABOX}
\end{figure}

Firstly, we generate ARMA DGP with the change point at $\theta=0.2$ and 0.4, and take the sample size $n=200, 500$ and  $1000$. We replicate 500 experiments for each case.
The Bias and ABias of the estimators are shown in Table \ref{ARMATABLE}. For saving space and because of similarity, we select the case  when $n=500$ with the model (\ref{model1}) and the model (\ref{model2}) ($\Delta_{\varphi}=0.5$), respectively, and plot the box-plots of the change point estimates in Figure \ref{ARMABOX}. Note that the estimates outside $1.5\cdot {\sf IQR}$ from the median are marked by the void circle as outliers, where ${\sf IQR}$ means the interquartile range.

We can see from the table that ``nwss" performs best among all cases and ``nwss" estimator has far smaller ``Bias'' and ``ABias'' than other three methods. When the change point is near time boundary ($\theta=0.2$), the performance is still not bad as long as the structural change or sample size is not too small, despite it is inferior to that when the change point is about the central position.
Intuitively, the local linear estimation may be superior to the N-W estimation.
However, the result is opposite when they are introduced in the change-point detection procedure. Large ``Bias'', ``BiasSd'',``ABias'' and ``ABiasSd'' are witnessed in most cases. We may call this phenomenon as the local linear paradox.

It is well known that the local linear fit is less influenced by outliers than the N-W fit, that is, the N-W estimation is more sensitive to outliers than the local linear estimation. Thus, we suspect that, when we use the local linear estimation,  $|\widehat\varphi_{1,t}-\widehat\varphi_{t+1,n}|$ is flat around $t=k$ and $W_{1,n}(t)$ probably reaches the maximum at other time point when some stochastic errors exist. Thus, we will get the wrong estimate of the change point. While using N-W estimation, $|\widehat\varphi_{1,t}-\widehat\varphi_{t+1,n}|$ decreases quickly at around $t=k$ because of its sensitivity property, and we can obtain the maximum at the change point $k$. The stability advantage of the local linear fit develops disadvantage in this change-point detection and the local linear regression cannot be applied well to our detection in practice.
\begin{table}[H]
\centering
\scalebox{0.8}{
\begin{tabular}{c|c|l|rrrr|rrrr}
\toprule
\multicolumn{1}{c|}{\multirow{2}*{n}}&
\multicolumn{1}{c|}{\multirow{2}*{Model}}&
\multicolumn{1}{c|}{\multirow{2}*{Method}}&
\multicolumn{4}{c|}{$\theta=0.4$}&
\multicolumn{4}{c}{$\theta=0.2$}\\
\cline{4-11}
\multicolumn{1}{c|}{}&
\multicolumn{1}{c|}{}&
\multicolumn{1}{c|}{}&  Bias& BiasSd& ABias& ABiasSd&Bias& BiasSd& ABias& ABiasSd\\
\cline{1-11}
\multicolumn{1}{c|}{\multirow{12}*{n=200}}&\multicolumn{1}{c|}{\multirow{4}*{4.1}}& nwss&2.126&7.233&2.802&6.998&5.278&14.034&5.938&13.767\\
\multicolumn{1}{c|}{}&\multicolumn{1}{c|}{}&nwsup&9.406&8.813&9.570&8.634&22.110&22.453&22.158&22.406\\
\multicolumn{1}{c|}{}&\multicolumn{1}{c|}{}&llsup&14.020&58.129&43.864&40.595&42.094&66.791&55.826&55.800\\
\multicolumn{1}{c|}{}&\multicolumn{1}{c|}{}&llss&12.686&68.348&51.760&46.343&42.170&76.537&60.950&62.590\\
\cline{2-11}
\multicolumn{1}{c|}{}&\multicolumn{1}{c|}{\multirow{4}*{4.2}}& nwss&2.013&16.465&8.963&13.951&8.313&22.781&12.722&20.638\\
\multicolumn{1}{c|}{}&\multicolumn{1}{c|}{\multirow{4}*{($\Delta_{\varphi}=0.5$)}}&nwsup&9.646&17.357&14.086&13.988&28.978&28.889&30.434&27.347\\
\multicolumn{1}{c|}{}&\multicolumn{1}{c|}{}&llsup&14.106&68.767&57.806&39.750&47.644&74.360&66.068&58.568\\
\multicolumn{1}{c|}{}&\multicolumn{1}{c|}{}&llss&14.362&74.272&61.334&44.201&43.322&80.185&65.522&63.313\\
\cline{2-11}
\multicolumn{1}{c|}{}&\multicolumn{1}{c|}{\multirow{4}*{4.2}}& nwss&2.0678&34.870&23.277&26.015&21.871&47.352&30.505&42.289\\
\multicolumn{1}{c|}{}&\multicolumn{1}{c|}{\multirow{4}*{($\Delta_{\varphi}=0.3$)}}&nwsup&11.788&27.759&21.620&21.011&41.736&39.748&43.928&37.307\\
\multicolumn{1}{c|}{}&\multicolumn{1}{c|}{}&llsup&16.316&76.542&69.908&35.050&53.682&79.217&74.818&59.615\\
\multicolumn{1}{c|}{}&\multicolumn{1}{c|}{}&llss&14.042&82.441&75.686&35.416&51.082&85.440&77.166&62.834\\
\cline{1-11}
\multicolumn{1}{c|}{\multirow{12}*{n=500}}&\multicolumn{1}{c|}{\multirow{4}*{4.1}}& nwss&1.192&3.463&1.788&3.196&2.728&6.877&3.236&6.653\\
\multicolumn{1}{c|}{}&\multicolumn{1}{c|}{}&nwsup&12.228&12.586&12.236&12.578&30.756&30.899&30.768&30.887\\
\multicolumn{1}{c|}{}&\multicolumn{1}{c|}{}&llsup&24.812&109.096&60.748&93.921&73.770&141.537&95.658&127.737\\
\multicolumn{1}{c|}{}&\multicolumn{1}{c|}{}&llss&16.688&158.394&104.268&120.308&77.944&183.979&126.028&154.986\\
\cline{2-11}
\multicolumn{1}{c|}{}&\multicolumn{1}{c|}{\multirow{4}*{4.2}}& nwss&3.098&12.404&6.889&10.767&6.845&18.480&10.949&16.381\\
\multicolumn{1}{c|}{}&\multicolumn{1}{c|}{\multirow{4}*{($\Delta_{\varphi}=0.5$)}}&nwsup&17.996&26.223&21.656&23.286&43.352&46.817&45.264&44.968\\
\multicolumn{1}{c|}{}&\multicolumn{1}{c|}{}&llsup&25.834&146.082&99.294&110.134&93.622&176.052&135.626&146.102\\
\multicolumn{1}{c|}{}&\multicolumn{1}{c|}{}&llss&21.488&174.659&126.644&122.056&87.646&197.860&144.898&160.649\\
\cline{2-11}
\multicolumn{1}{c|}{}&\multicolumn{1}{c|}{\multirow{4}*{4.2}}& nwss&4.359&46.524&24.038&40.055&14.403&69.204&32.088&62.967\\
\multicolumn{1}{c|}{}&\multicolumn{1}{c|}{\multirow{4}*{($\Delta_{\varphi}=0.3$)}}&nwsup&19.922&46.556&35.034&36.542&71.740&79.929&76.080&75.801\\
\multicolumn{1}{c|}{}&\multicolumn{1}{c|}{}&llsup&36.958&180.757&146.014&112.601&115.930&198.597&172.182&152.320\\
\multicolumn{1}{c|}{}&\multicolumn{1}{c|}{}&llss&27.194&204.130&173.426&110.786&108.079&217.664&182.138&160.752\\
\cline{1-11}
\multicolumn{1}{c|}{\multirow{12}*{n=1000}}&\multicolumn{1}{c|}{\multirow{4}*{4.1}}& nwss&1.732&4.743&2.280&4.505&2.188&5.593&2.504&5.462\\
\multicolumn{1}{c|}{}&\multicolumn{1}{c|}{}&nwsup&15.914&18.122&15.930&18.108&42.434&45.790&42.442&45.783\\
\multicolumn{1}{c|}{}&\multicolumn{1}{c|}{}&llsup&31.280&159.419&67.836&147.593&94.096&221.589&121.688&207.691\\
\multicolumn{1}{c|}{}&\multicolumn{1}{c|}{}&llss&41.962&282.981&160.130&236.959&132.434&342.388&202.526&306.112\\
\cline{2-11}
\multicolumn{1}{c|}{}&\multicolumn{1}{c|}{\multirow{4}*{4.2}}&nwss&3.464&9.112&5.616&7.965&6.730&16.901&9.570&15.469\\
\multicolumn{1}{c|}{}&\multicolumn{1}{c|}{\multirow{4}*{($\Delta_{\varphi}=0.5$)}}&nwsup&25.676&35.052&28.544&32.754&61.428&72.720&62.800&71.536\\
\multicolumn{1}{c|}{}&\multicolumn{1}{c|}{}&llsup&37.054&228.565&119.742&198.119&137.230&305.007&191.638&274.044\\
\multicolumn{1}{c|}{}&\multicolumn{1}{c|}{}&llss&45.342&312.549&196.314&247.246&148.852&373.300&243.264&319.779\\
\cline{2-11}
\multicolumn{1}{c|}{}&\multicolumn{1}{c|}{\multirow{4}*{4.2}}& nwss&4.404&33.797&16.152&30.005&6.615&38.622&22.373&32.154\\
\multicolumn{1}{c|}{}&\multicolumn{1}{c|}{\multirow{4}*{($\Delta_{\varphi}=0.3$)}}&nwsup&32.278&59.658&46.874&49.005&96.584&113.546&100.448&110.136\\
\multicolumn{1}{c|}{}&\multicolumn{1}{c|}{}&llsup&62.740&305.363&203.232&236.230&175.600&360.074&262.188&302.771\\
\multicolumn{1}{c|}{}&\multicolumn{1}{c|}{}&llss&64.268&384.523&299.620&249.088&202.762&425.231&332.922&333.103\\
\bottomrule
\end{tabular}}
\caption{\ Simulating $X_t$ and $\epsilon_t$ by ARMA DGP, we replicate enormous experiments under different sample sizes, models and amounts of structural change. Then we compute the estimation bias of each method. }\label{ARMATABLE}
\end{table}

\begin{figure}[H]
\centering
\subfigure[$\theta=0.4,n=500$, and ARFIMA DGP under model (4.1)]{\includegraphics[width=0.35\textwidth]{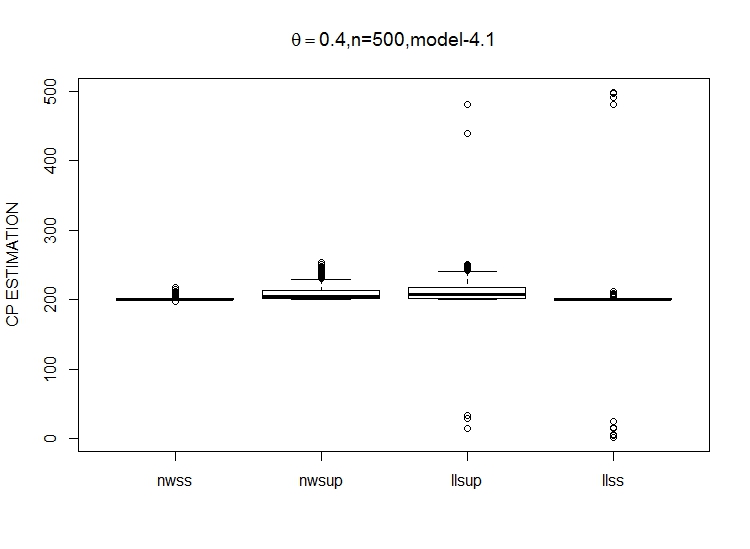}}\\
\subfigure[$\theta=0.4,n=500$, and ARFIMA DGP under model (4.2)]{\includegraphics[width=0.35\textwidth]{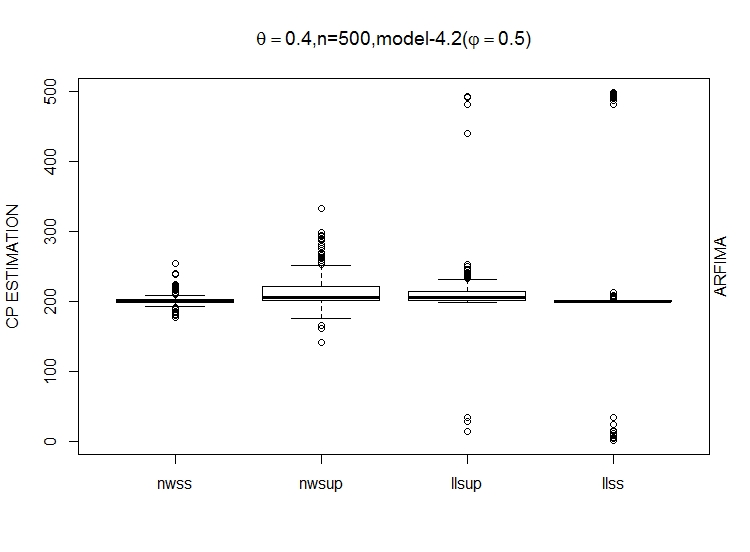}}\\
\subfigure[$\theta=0.2,n=500$, and ARFIMA DGP under model (4.1)]{\includegraphics[width=0.35\textwidth]{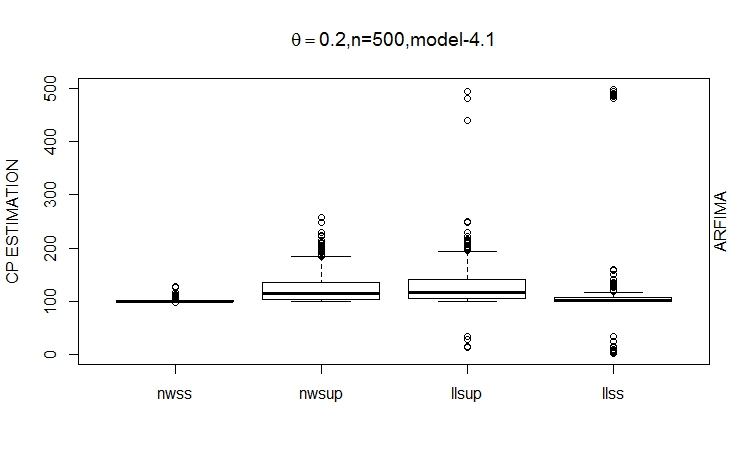}}\\
\subfigure[$\theta=0.2,n=500$, and ARFIMA DGP under model (4.2)]{\includegraphics[width=0.35\textwidth]{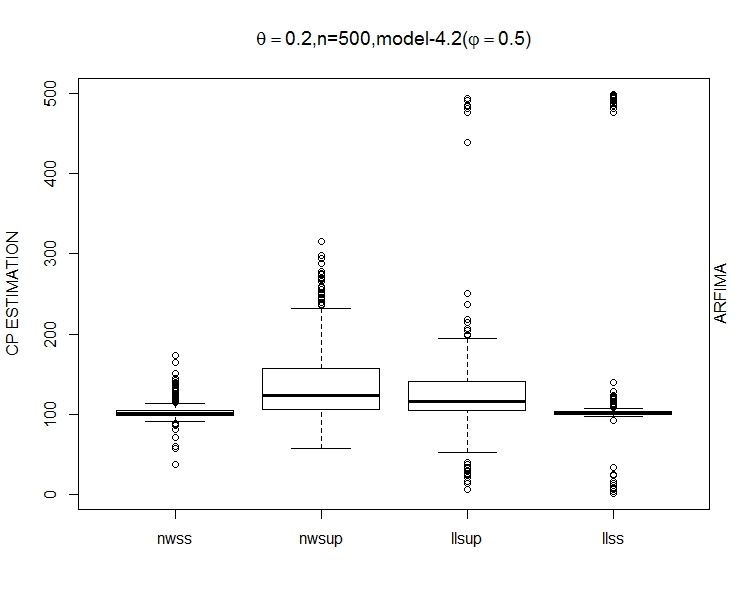}}
\caption{\ At the case when data is generated by ARFIMA DGP, setting sample size $n=500$, we plot the box-plots of the change point estimates under the model (\ref{model1}) and the model (\ref{model2}) with $\theta=0.4$ and $\theta=0.2$, respectively.}\label{ARFIMABOX}
\end{figure}

To compare our method ``nwss" with  ``nwsup" in \cite{np1}, we consider the long-memory DGP given by (\ref{eq.4.4}) and (\ref{4.5}).
The results are shown in Table \ref{ARFIMATABLE} and Figure \ref{ARFIMABOX}. Even for the long-memory data, ``nwss" still has a superb performance with few big deviations as long as the structural change is not too tiny. While the method ``nwsup" has relatively big bias and standard deviation. Also, the local linear paradox remains exist.

\begin{table}[H]
\centering
\scalebox{0.9}{
\begin{tabular}{c|c|l|rrrr|rrrr}
\toprule
\multicolumn{1}{c|}{\multirow{2}*{n}}&
\multicolumn{1}{c|}{\multirow{2}*{Model}}&
\multicolumn{1}{c|}{\multirow{2}*{Method}}&
\multicolumn{4}{c|}{$\theta=0.4$}&
\multicolumn{4}{c}{$\theta=0.2$}\\
\cline{4-11}
\multicolumn{1}{c|}{}&
\multicolumn{1}{c|}{}&
\multicolumn{1}{c|}{}&  Bias& BiasSd& ABias& ABiasSd&Bias& BiasSd& ABias& ABiasSd\\
\cline{1-11}
\multicolumn{1}{c|}{\multirow{12}*{n=200}}&\multicolumn{1}{c|}{\multirow{4}*{4.1}}& nwss&0.786&1.870&0.858&1.838&1.745&4.203&1.806&4.177\\
\multicolumn{1}{c|}{}&\multicolumn{1}{c|}{}&nwsup&6.698&7.370&6.702&7.367&16.330&19.211&16.334&19.207\\
\multicolumn{1}{c|}{}&\multicolumn{1}{c|}{}&llsup&7.146&16.438&9.794&15.009&17.624&26.318&19.608&24.844\\
\multicolumn{1}{c|}{}&\multicolumn{1}{c|}{}&llss&0.700&21.658&6.056&20.804&8.528&30.962&11.856&29.844\\
\cline{2-11}
\multicolumn{1}{c|}{}&\multicolumn{1}{c|}{\multirow{4}*{4.2}}& nwss&1.673&10.756&4.861&9.737&6.807&22.572&9.718&21.475\\
\multicolumn{1}{c|}{}&\multicolumn{1}{c|}{\multirow{4}*{($\Delta_{\varphi}=0.5$)}}&nwsup&7.106&14.642&10.890&12.090&23.486&27.060&24.798&25.860\\
\multicolumn{1}{c|}{}&\multicolumn{1}{c|}{}&llsup&7.448&26.129&13.888&23.346&21.586&36.812&25.422&34.270\\
\multicolumn{1}{c|}{}&\multicolumn{1}{c|}{}&llss&1.394&28.162&9.482&26.551&7.820&34.338&12.140&33.056\\
\cline{2-11}
\multicolumn{1}{c|}{}&\multicolumn{1}{c|}{\multirow{4}*{4.2}}& nwss&4.227&40.089&23.409&32.693&23.933&53.243&32.733&48.216\\
\multicolumn{1}{c|}{}&\multicolumn{1}{c|}{\multirow{4}*{($\Delta_{\varphi}=0.3$)}}&nwsup&9.984&25.850&18.652&20.482&33.624&36.117&35.928&33.821\\
\multicolumn{1}{c|}{}&\multicolumn{1}{c|}{}&llsup&10.890&39.516&23.346&33.679&28.224&48.451&35.040&43.766\\
\multicolumn{1}{c|}{}&\multicolumn{1}{c|}{}&llss&5.010&41.815&19.53&37.298&16.232&51.461&24.752&47.941\\
\cline{1-11}
\multicolumn{1}{c|}{\multirow{12}*{n=500}}&\multicolumn{1}{c|}{\multirow{4}*{4.1}}& nwss&0.922&1.999&0.954&1.984&1.162&2.728&1.182&2.719\\
\multicolumn{1}{c|}{}&\multicolumn{1}{c|}{}&nwsup&8.904&9.966&8.904&9.966&24.362&27.943&24.370&27.936\\
\multicolumn{1}{c|}{}&\multicolumn{1}{c|}{}&llsup&11.076&24.188&13.168&23.114&28.236&39.803&29.464&38.901\\
\multicolumn{1}{c|}{}&\multicolumn{1}{c|}{}&llss&0.696&37.550&6.708&36.952&11.318&59.164&15.502&58.206\\
\cline{2-11}
\multicolumn{1}{c|}{}&\multicolumn{1}{c|}{\multirow{4}*{4.2}}& nwss&1.506&6.405&3.730&5.419&3.544&10.613&5.792&9.572\\
\multicolumn{1}{c|}{}&\multicolumn{1}{c|}{\multirow{4}*{($\Delta_{\varphi}=0.5$)}}&nwsup&13.208&21.384&15.984&19.393&38.590&44.179&39.530&43.338\\
\multicolumn{1}{c|}{}&\multicolumn{1}{c|}{}&llsup&10.114&29.557&12.210&28.753&27.186&52.429&32.262&49.462\\
\multicolumn{1}{c|}{}&\multicolumn{1}{c|}{}&llss&2.256&52.782&12.024&51.441&13.924&76.997&20.320&75.558\\
\cline{2-11}
\multicolumn{1}{c|}{}&\multicolumn{1}{c|}{\multirow{4}*{4.2}}& nwss&2.184&25.692&11.762&22.939&5.550&41.453&16.631&38.258\\
\multicolumn{1}{c|}{}&\multicolumn{1}{c|}{\multirow{4}*{($\Delta_{\varphi}=0.3$)}}&nwsup&18.374&34.478&26.066&29.090&58.748&66.250&62.484&62.731\\
\multicolumn{1}{c|}{}&\multicolumn{1}{c|}{}&llsup&8.156&53.917&21.940&49.914&39.048&83.426&49.576&77.620\\
\multicolumn{1}{c|}{}&\multicolumn{1}{c|}{}&llss&6.146&78.835&25.726&74.764&25.374&107.586&38.206&103.717\\
\cline{1-11}
\multicolumn{1}{c|}{\multirow{12}*{n=1000}}&\multicolumn{1}{c|}{\multirow{4}*{4.1}}& nwss&0.921&1.998&0.938&1.991&1.346&2.934&1.354&2.930\\
\multicolumn{1}{c|}{}&\multicolumn{1}{c|}{}&nwsup&11.548&13.829&11.548&13.829&33.014&38.612&33.014&38.612\\
\multicolumn{1}{c|}{}&\multicolumn{1}{c|}{}&llsup&15.088&28.843&15.088&28.843&39.944&53.390&40.628&52.871\\
\multicolumn{1}{c|}{}&\multicolumn{1}{c|}{}&llss&1.368&56.758&7.500&56.276&15.296&101.836&20.520&100.912\\
\cline{2-11}
\multicolumn{1}{c|}{}&\multicolumn{1}{c|}{\multirow{4}*{4.2}}& nwss&1.198&5.542&3.138&4.721&2.898&9.249&5.006&8.298\\
\multicolumn{1}{c|}{}&\multicolumn{1}{c|}{\multirow{4}*{($\Delta_{\varphi}=0.5$)}}&nwsup&14.156&23.915&16.780&22.149&54.828&64.433&55.952&63.457\\
\multicolumn{1}{c|}{}&\multicolumn{1}{c|}{}&llsup&9.540&36.459&12.492&35.554&39.848&64.470&41.912&63.145\\
\multicolumn{1}{c|}{}&\multicolumn{1}{c|}{}&llss&2.424&79.112&13.268&78.027&13.556&114.871&22.468&113.462\\
\cline{2-11}
\multicolumn{1}{c|}{}&\multicolumn{1}{c|}{\multirow{4}*{4.2}}& nwss&2.610&29.752&8.870&28.516&5.829&45.781&13.036&44.268\\
\multicolumn{1}{c|}{}&\multicolumn{1}{c|}{\multirow{4}*{($\Delta_{\varphi}=0.3$)}}&nwsup&17.814&44.65646&29.966&37.582&75.358&90.264&77.698&88.253\\
\multicolumn{1}{c|}{}&\multicolumn{1}{c|}{}&llsup&10.604&60.211&18.692&58.206&51.374&102.357&57.290&99.161\\
\multicolumn{1}{c|}{}&\multicolumn{1}{c|}{}&llss&9.254&127.826&32.570&123.945&34.000&177.884&49.164&174.296\\
\bottomrule
\end{tabular}}
\caption{\ Simulating $X_t$ and $\epsilon_t$ by ARFIMA DGP, we replicate enormous experiments under different sample sizes, models and amounts of structural change. Then we compute the estimation bias of each method.}\label{ARFIMATABLE}
\end{table}

In conclusion, on one hand, our method is superior to other methods in each case and it performs well as long as the structural change is not too tiny and the change point is not too close to the edge, and the robust application to the long-memory data has been shown by ARFIMA DGP. On the other hand, ``nwsup" tends to have a relatively bad performance. Last but not least, ``llss'' does not promote the effect and even results in big deviations in most cases, which pushes us only to use the simpler  N-W estimation.

\section{Real data analysis}\label{sec5}

We apply our method to the Nasdaq 100 index for finding the structural change between the return volatility and the return. The relation between volatility and return is sometimes referred to as leverage effect, which can be illustrated as an asymmetric U-shaped curve. This relation can be time-varying, see for example \cite{BR12} and \cite{Jin17}.

\begin{figure}[H]
\centering
\subfigure[The sequence of volatility $Y_t$]{\includegraphics[width=0.6\textwidth]{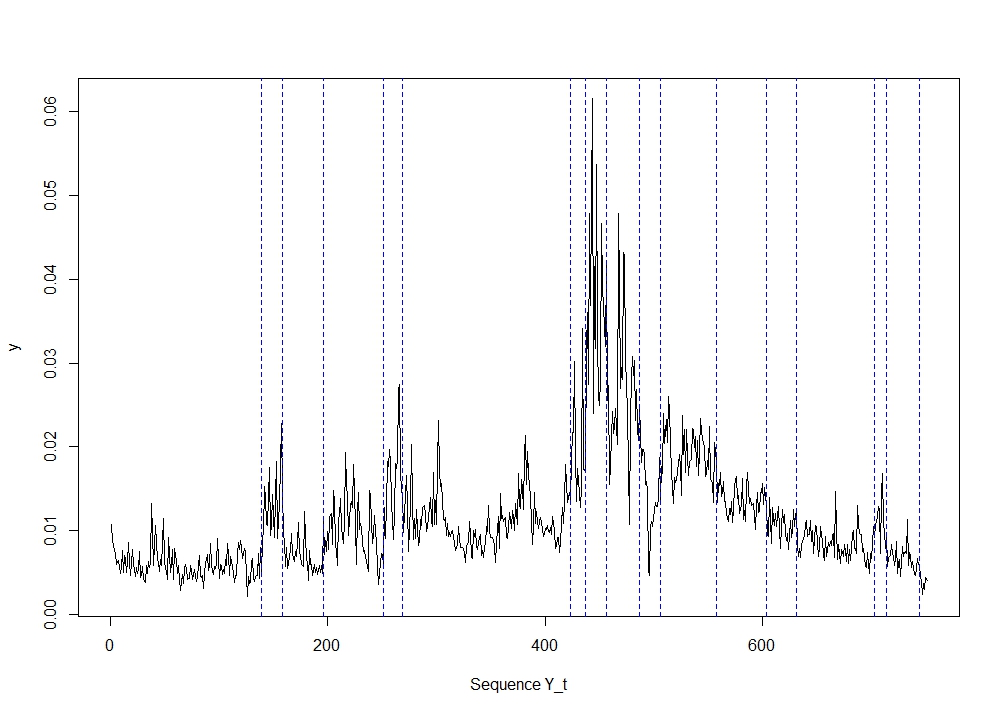}}\\
\subfigure[The sequence of return $X_t$]{\includegraphics[width=0.6\textwidth]{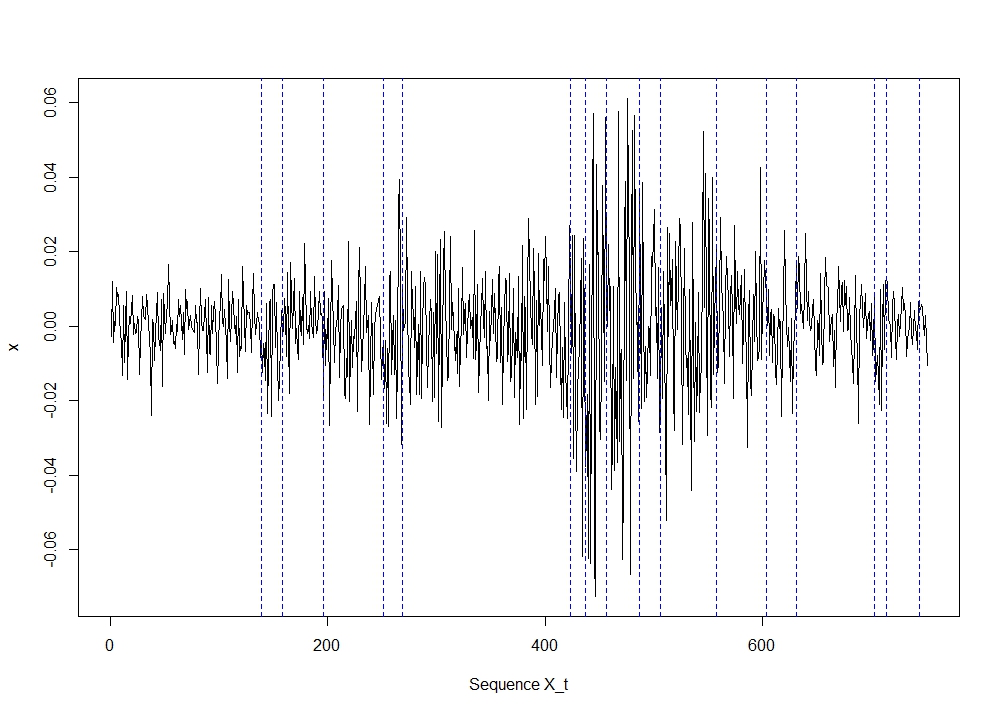}}
\caption{\ The sequences of return $X_t$ and  volatility $Y_t$}\label{realdataxy}
\end{figure}

%It is widely agreed that, although daily and monthly financial asset returns are approximately  unpredictable, return volatility is highly predictable. Of course,
The volatility is inherently unobservable and we consider the realized volatility as the proxy of the return volatility.
A discussion in gauging return-volatility regressions using different volatility measures can be found in \cite{BZ06}.
We take the three-year Nasdaq 100 index data from 2007--2009 as the sample (reader can acquire the data from the site \url {https://realized.oxford-man.ox.ac.uk/data}), and set its realized volatility as $Y_t$ and return as $X_t$.
We plot the sequence $Y_t$ and $X_t$ in Figure  \ref{realdataxy}, and think that there probably exists multiple change points. The dotted vertical lines mark the change-point positions computed later.

\begin{figure}[H]
  \centering
  \includegraphics[width=0.8\textwidth]{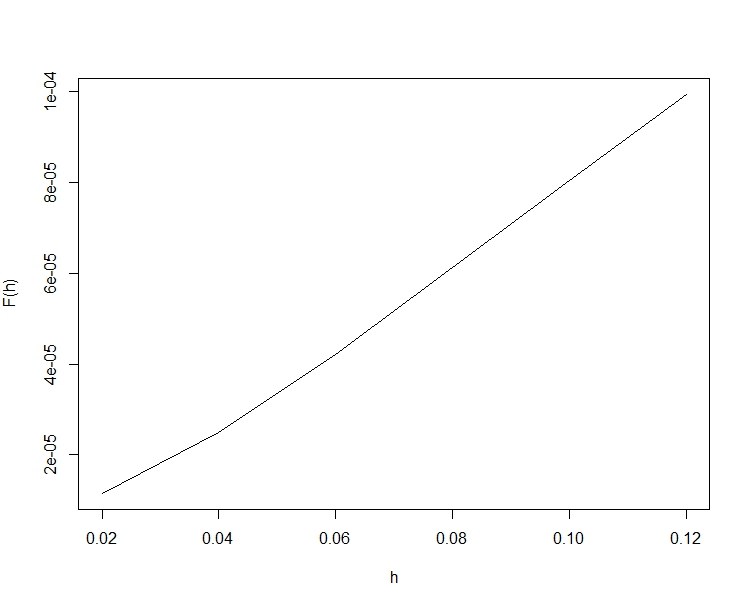}
    \caption{\ $F(h)$ based on Nasdaq index from 2007 to 2009}\label{Fhrealdata}
\end{figure}

In order to select the bandwidth, we plot the curve $F(h)$ on the interval $(0,\max(X_t)-\min(X_t))$ in Figure \ref{Fhrealdata}. It can be seen that $F(h)$ increases monotonically, so we choose  the bandwidth $h=0.12$.  Then we obtain the threshold by the permutation method as proposed in the simulation.

We detect the change point from the whole sample, and obtain the change point at 12/31/2007. Then we split the original sample to two sub-samples. Repeating the same steps for the left and right sub-samples, we detect the second change point 07/23/2007 and third one 06/02/2009. Note that the threshold is always changing because of the change of the sample size. Repeating the binary segmentation until the maximum of $W_{1,n}(t)$ of each sub-sample is less than the threshold or the sub-sample sizes are too few, we totally detect 16 change points (dates). We investigate the corresponding big financial events resulting in stock  market to fluctuate just near the date in Table \ref{realdatatable}.
%It can be seen that, although the variance of $X_t$ changes dramatically once, $X_t$ keeps stable approximately in the local range. However, in the local range, $Y_t$ often changes dramatically, which is probably caused by the structural change of $\varphi(x)$.  In fact, they appeared during the global financial crisis.
We can expect that during the financial crisis, prices and volatility will be unstable. From our analysis, we can see that the relationship between price and volatility is also fragile, and change points occur very frequently.

\begin{figure}[H]
\centering
\includegraphics[width=0.9\textwidth]{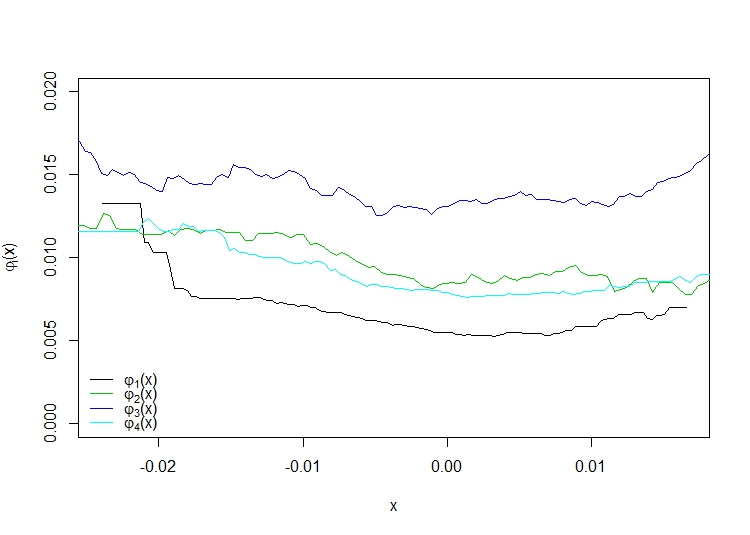}
\caption{\ N-W kernel regression curves: $\varphi_1(x)$ from 01/03/2007 to 07/23/2007, $\varphi_2(x)$ from 07/23/2007 to 12/31/2007, $\varphi_3(x)$ from 12/31/2007 to 06/02/2009, $\varphi_4(x)$ from 06/02/2009 to 12/31/2009, respectively.}\label{realdatavarphi}
\end{figure}

To see how the leverage curve changes visually, we choose the first three detected change points 07/23/2007, 12/31/2007 and 06/02/2009 as the split points, and plot the four corresponding classical N-W kernel regression curves in Figure \ref{realdatavarphi} with the bandwidth $h=0.01$. % Note that now we seek for the perfect fit of regression curves and not use the detection bandwidth.
It is clear to find the obvious differences between $\widehat\varphi_{1}$ and $\widehat\varphi_{2}$, $\widehat\varphi_{2}$ and $\widehat\varphi_{3}$, $\widehat\varphi_{3}$ and $\widehat\varphi_{4}$, respectively.
%, we think our method is not bad when it is applied in the real data.

\begin{table}[H]
\centering
\scalebox{0.8}{
\begin{tabular}{c|c|c|c}
\toprule
date (change point)& event &date (change point)&event\\
\cline{1-4}

\multicolumn{1}{c|}{\multirow{2}*{07/23/2007}}&Largest US commercial mortgage company&
\multicolumn{1}{c|}{\multirow{2}*{08/17/2007}}&American FED cut the window discount\\
\multicolumn{1}{c|}{}&declared profit decreased by 33\%&
\multicolumn{1}{c|}{}& rate by 50 basis points to 5.75\\
\cline{1-4}

\multicolumn{1}{c|}{\multirow{2}*{10/10/2007}}&The dow Jones closed at &
\multicolumn{1}{c|}{\multirow{2}*{12/31/2007}}&International agricultural futures  \\
\multicolumn{1}{c|}{}& an all-time high of 14,165&
\multicolumn{1}{c|}{}&prices continue to set new records\\
\cline{1-4}

\multicolumn{1}{c|}{\multirow{2}*{01/28/2008}}&American FED provided \$30 billion&
\multicolumn{1}{c|}{\multirow{2}*{09/12/2008}}&The 4th large US investment\\
\multicolumn{1}{c|}{}&to commercial bank&
\multicolumn{1}{c|}{}&bank filed bankruptcy protection\\%American sub-prime mortgage crisis developed to global financial crisis.\\
\cline{1-4}

\multicolumn{1}{c|}{\multirow{2}*{10/02/2008}}&Bush signed a \$700&
\multicolumn{1}{c|}{\multirow{2}*{10/29/2008}}&American sub-prime mortgage crisis\\
\multicolumn{1}{c|}{}&billion financial rescue plan&
\multicolumn{1}{c|}{}& developed to global financial crisis\\
\cline{1-4}

\multicolumn{1}{c|}{\multirow{2}*{12/12/2008}}&Many Banks in world cut&
\multicolumn{1}{c|}{\multirow{2}*{01/12/2009}}&The Finance of America  injected \\
\multicolumn{1}{c|}{}&interest rates in tandem once more&
\multicolumn{1}{c|}{}&\$14.77 billion into 43 Banks\\
\cline{1-4}

\multicolumn{1}{c|}{\multirow{2}*{03/26/2009}}&American FED declared to use \$200 billion&
\multicolumn{1}{c|}{\multirow{2}*{06/02/2009}}&General motors officially filed for \\
\multicolumn{1}{c|}{}&to help personal consumers and small enterprises&
\multicolumn{1}{c|}{}& bankruptcy\\
\cline{1-4}

\multicolumn{1}{c|}{\multirow{2}*{07/13/2009}}&England bank announced it would&
\multicolumn{1}{c|}{\multirow{2}*{10/22/2009}}&The European debt crisis\\
\multicolumn{1}{c|}{}&maintain QE at the 120 billion pounds  &
\multicolumn{1}{c|}{}&officially erupted\\
\cline{1-4}

\multicolumn{1}{c|}{\multirow{2}*{11/06/2009}}&England bank announced it would&
\multicolumn{1}{c|}{\multirow{2}*{12/21/2009}}&US House of Representatives passed the biggest \\
\multicolumn{1}{c|}{}&increase QE 20 billion to 200 billion pounds&
\multicolumn{1}{c|}{}& financial regulatory reform since 1930s\\
\bottomrule
\end{tabular}}
\caption{\ The probable change points and the big financial events around them} \label{realdatatable}
\end{table}

\section{Conclusion and discussion}\label{sec6}
In this paper, firstly, we construct a CUSUM statistic to detect the structural change in the nonparametric regression model. The CUSUM statistic is constructed using sum-of-squares but not supremum like \cite{np1} so that we can find the relatively small structural change, especially when the function difference is small in the whole domain.
Secondly, with probability one, we derive an upper bound of the CUSUM statistic asymptotically when there is no change point as well as a lower bound asymptotically when there is a change point. We establish the strong consistency of the change point estimator. Although we detect the change point by the threshold method not usual hypothesis testing, we still show in the simulation that our method has a low rate of the false positive and high rate of the true positive even if the structural change of the regression function is not obvious. Last but not least, we demonstrate a surprising result that the CUSUM method constructed by the N-W estimator performs better than that constructed by the local linear estimator, because the N-W estimator is more sensitive to the observations or outliers which do not belong to the same stable period.

Although we only focus on the case of one regressor and one change point, our method can be extended without difficulties to the multivariate regression with multiple change points.

\begin{acknowledgements}
The authors thank the professor Cai-Ya Zhang from ZUCC, senior brothers and many schoolmates for the vital comments and suggestions. Specially, we appreciate the editor and reviewers for their comments and suggestions to our research, which improve our work significantly.
\end{acknowledgements}

\section*{Appendix: Proofs}\label{Appendix}
\renewcommand{\theequation}{6.\arabic{equation}}
\setcounter{equation}{0}

In this section, we provide the detailed proofs of the theoretical results in Section \ref{sec3}. Before proving the main theorems, we state and prove some lemmas.

The following lemma plays a crucial role in deriving some uniform bounds of an $\alpha$-mixing process.

\begin{lemma}[Theorem 1.3 of \cite{Bo98}]\label{ppst:mixing}
Let $(X_t , t\in \mathbb{Z})$ be a zero-mean real-valued process such that
$\sup \limits_{1\leq t\leq n}\Vert X_t\Vert_\infty\leq b$. Let~$S_n=\sum_{t=1}^nX_t$. Then\\
(i) For each integer $q\in[1,\frac{n}{2}]$ and each $\varepsilon>0$,
$${\sf P}(|S_n|>n\varepsilon)\leq 4\exp\left(-\frac{\varepsilon^2}{8b^2}q\right)
+22\left(1+\frac{4b}{\varepsilon}\right)^{1/2}q\alpha\left(\left\lfloor\frac{n}{2q}\right\rfloor\right).$$
(ii) For each integer $q\in[1,\frac{n}{2}]$ and each $\varepsilon>0$,
$${\sf P}(|S_n|>n\varepsilon)\leq 4\exp\left(-\frac{\varepsilon^2}{8v^2(q)}q\right)
+22\left(1+\frac{4b}{\varepsilon}\right)^{1/2}q\alpha\left(\left\lfloor\frac{n}{2q}\right\rfloor\right)$$
with $v^2(q)=\frac{2}{p^2}\sigma^2(q)+\frac{b\varepsilon}{2}$,
$p=\frac{n}{2q}$,
\begin{eqnarray*}
\sigma^2(q)&=&\max_{0\leq j\leq 2q-1}{\sf E}\left\{(\lfloor jp\rfloor+1-jp)X_{\lfloor jp\rfloor+1}+X_{\lfloor jp\rfloor+2}+\right.\notag\\
&&\left.\cdots+X_{\lfloor(j+1)p\rfloor}+((j+1)p-\lfloor(j+1)p\rfloor)X_{\lfloor(j+1)p+1\rfloor}\right\} ^{2}.
\end{eqnarray*}
\end{lemma}

\smallskip

The following lemma demonstrates the bounds of the auto-covariances of $K_{h_n}(x-X_i)$ and $Y_i I_{(|Y_i|\leq T)} K_{h_n}(x-X_i)$, that is, a truncated version of $Y_iK_{h_n}(x-X_i)$, which facilitates the application of Lemma 1 to the N-W estimator.

\begin{lemma}\label{lemma2}
Suppose $\{X_t,Y_t\}_{t=1}^n$ is an $\alpha$-mixing process, and Assumptions \ref{a4}--\ref{a6} are satisfied. Then we have\\
(i)
\begin{equation}\label{eq6.2}
{\sf Var}\left(K_{h_n}(x-X_i)\right)\leq c_1 h_n^{-1},
\end{equation}
and
\begin{equation}\label{eq6.3}
\left|{\sf Cov}\left(K_{h_n}(x-X_i),K_{h_n}(x-X_j)\right)\right|
\leq  \left\{
\begin{aligned}
c_2\min \{h_n^{-1+q_F},h_n^{-2}|i-j|^{-\gamma}\}, \{X_t,Y_t\}_{t=1}^n \in PSM,\\
c_2\min \{h_n^{-1+q_F},h_n^{-2}\rho^{|i-j|}\}\ \ \ \ , \{X_t,Y_t\}_{t=1}^n \in GSM\\
\end{aligned}
\right.
\end{equation}
for any $1\leq i\neq j\leq n$  and $x \in \mathbb{R}$,
where $c_1$ and $c_2$ are two positive constants, and $q_F$ is defined in Assumption \ref{a5}.\\
(ii)
\begin{equation}\label{eq6.4}
{\sf Var}(Y_i I_{(|Y_i|\leq T)} K_{h_n}(x-X_i))\leq c_3 h_n^{-1},
\end{equation}
and
\begin{eqnarray}\label{eq6.5}
&&\left|{\sf Cov}\left(Y_i I_{(|Y_i|\leq T)} K_{h_n}(x-X_i),Y_j I_{(|Y_j|\leq T)} K_{h_n}(x-X_j)\right)\right|\notag\\
&\leq& \left\{
\begin{aligned}
c_4\min \{h_n^{-1+q_F},h_n^{-2}|i-j|^{-\gamma}\}, \  \{X_t,Y_t\}_{t=1}^n \in PSM,\\
c_4\min \{h_n^{-1+q_F},h_n^{-2}\rho^{|i-j|}\}\ \ \ \ \ , \{X_t,Y_t\}_{t=1}^n \in GSM\\
\end{aligned}
\right.
\end{eqnarray}
for any $ 1\leq i\neq j\leq n$ and $x \in \mathbb{R}$, where $c_3$ and $c_4$ are two positive constants which does not dependent on $T$.
\end{lemma}
\smallskip

\noindent\textbf{Proof of Lemma \ref{lemma2}}

(i)  Noting that $f$, the density function of $X$, is uniformly bounded and $\|K\|_2<\infty$, we can prove that
\begin{eqnarray*}
{\sf Var}(K_{h_n}(x-X_i))&\leq&{\sf E}\left[K_{h_n}^2(x-X_i)\right]=\int_{\mathbb{R}} K_{h_n}^2(u-x)f(u)du\\
&=&h_n^{-1}\int_{\mathbb{R}} K^2(z)f(x+h_nz)dz
\leq c_1 h_n^{-1}
\end{eqnarray*}
with variable substitution $z=(u-x)/h_n$ and selecting $c_1=\Vert f\Vert_\infty\Vert K\Vert_2^2$.
In terms of the covariance, we have
\begin{eqnarray*}
&&{\sf Cov}\left(K_{h_n}(x-X_i),K_{h_n}(x-X_j)\right)\notag\\
&=&{\sf E}\left[K_{h_n}(x-X_i)K_{h_n}(x-X_j)\right]
-{\sf E}\left[K_{h_n}(x-X_i)\right]{\sf E}\left[K_{h_n}(x-X_j)\right]\notag\\
&=&\int_{{\mathbb{R}}^2} K_{h_n}(x-u)K_{h_n}(x-v)F^{(|i-j|)}(u,v)dudv,
\end{eqnarray*}
where $F^{(|i-j|)}$ is defined in Assumption \ref{a5}. Letting $\bar{p_F}$ satisfy $p_F^{-1}+\bar{p_F}^{-1}=1$ and using H\"{o}lder inequality, we can prove that
\begin{eqnarray}\label{eq6.6}
\left|{\sf Cov}\left(K_{h_n}(x-X_i),K_{h_n}(x-X_j)\right)\right|
\leq h_n^{-2}\cdot h_n^{2/\bar{p_F}}\|K\|_{\bar{p_F}}^2\cdot\|F^{|i-j|}\|_{p_F}\leq c_{2,1} h_n^{-1+q_F},
\end{eqnarray}
 where $c_{2,1}$ is equal to $\|K\|_{\bar{p_F}}^2\cdot C_2$, noting that $\|K\|_{\bar{p_F}}<\infty$, which is implied by $\|K\|_1<\infty$ and $\|K\|_\infty<\infty$ in Assumption \ref{a4}.
Besides, by using the Billingsley's inequality (c.f. Chapter 1 of \cite{Bo98}), we have
\begin{eqnarray}\label{eq6.7}
\left|{\sf Cov}\left(K_{h_n}(x-X_i),K_{h_n}(x-X_j)\right)\right|&\leq & 4\|K_{h_n}(x-X_i)\|_\infty\|K_{h_n}(x-X_j)\|_\infty \cdot\alpha(|i-j|)\notag\\
&\leq& c_{2,2}h_n^{-2}\alpha(|i-j|),
\end{eqnarray}
where $c_{2,2}=4\|K\|_\infty^2$.
Note that we have $\alpha(|i-j|)\leq C_1\rho^{|i-j|}$ in Assumption \ref{a2} for GSM or $\alpha(|i-j|)\leq C_1|i-j|^{-\gamma}$ in  Assumption \ref{a1} for PSM.
Combining (\ref{eq6.6}) and (\ref{eq6.7}), and
taking $c_2=\max\{c_{2,1},c_{2,2}C_1\}$, we can prove (\ref{eq6.3}).
\smallskip

(ii) We denote $I_{(|Y_i|\leq T)}$ as $I_i$ for simplicity. Noting that $\|K\|_2<\infty$,
$\|f\|_\infty<\infty$ and $\sup_{x\in {\sf supp}[X_1]}{\sf E}\left[Y_i^2|X_i=x\right]<C_4$ by Assumptions \ref{a4}, \ref{a5} and \ref{a6}, respectively, we have
\begin{eqnarray*}
{\sf Var}(Y_iI_iK_{h_n}(x-X_i))\leq{\sf E}[Y_i^2K_{h_n}^2(x-X_i)]=\int_{\mathbb{R}} {\sf E}[Y_i^2|X_i=u]K_{h_n}^2(x-u)f(u)du\leq c_3 h_n^{-1}
\end{eqnarray*}
by selecting $c_3=C_4\cdot\Vert f\Vert_\infty\cdot\Vert K\Vert_2^2$.

For the covariance, set $A=\{(y_i,y_j):|y_i|\leq T,|y_j|\leq T\}$, we have
\begin{eqnarray*}\label{eq6.8}
&&\left|{\sf Cov}\left(Y_i I_iK_{h_n}(x-X_i),Y_jI_j K_{h_n}(x-X_j)\right)\right|\notag\\
&=&\left|{\sf E}\left[Y_i I_iK_{h_n}(x-X_i)\cdot Y_j I_jK_{h_n}(x-X_j)\right]-{\sf E}\left[Y_i I_iK_{h_n}(x-X_i)\right]
\cdot{\sf E}[Y_j I_jK_{h_n}(x-X_j)]\right|\notag\\
&=&\left|\int_{\mathbb{R}^2} \int_ {A} \left\{y_iy_jp(y_i,y_j,u,v)-y_ip(y_i,u)\cdot y_jp(y_j,v)\right\}dy_idy_j\cdot K_{h_n}(x-u)K_{h_n}(x-v)dudv\right|\notag\\
&\leq&\int_{\mathbb{R}^2} \int_ {\mathbb{R}^2} \left|y_iy_jp(y_i,y_j,u,v)-y_ip(y_i,u)y_jp(y_j,v)\right|dy_idy_j\cdot |K_{h_n}(x-u)K_{h_n}(x-v)|dudv\notag\\
&=&\int_{\mathbb{R}^2} G^{|i-j|}(u,v)\cdot\left|K_{h_n}(x-u)K_{h_n}(x-v)\right|dudv\notag\\
&\leq& h_n^{-2}\cdot h_n^{2/\bar{p_G}}\|K\|_{\bar{p_G}}^2\cdot\|G^{|i-j|}\|_{p_G}\leq c_{4,1} h_n^{-1+q_G},
\end{eqnarray*}
where $c_{4,1}=C_3\|K\|_{\bar{p_G}}^2$  and $C_3$ are defined in Assumption \ref{a5}. Note that the last step follows from the H\"{o}lder inequality similar to (\ref{eq6.6}) with $\bar{p_G}$ satisfying $p_G^{-1}+\bar{p_G}^{-1}=1$.

Next we prove the second part in the minimization  function in (\ref{eq6.5}). Note that, for any $m\geq1$ and $i=1,\cdots,n$,
\begin{eqnarray}\label{eq6.9}
 {\sf E}\left[|Y_iI_iK_{h_n}(X_i-x)|^m\right]
 &\leq& {\sf E}\left[|Y_iK_{h_n}(X_i-x)|^m\right]\notag\\
 &\leq& h_n^{-m}\cdot{\sf E}[|Y_i|^m]\cdot\Vert K\Vert_\infty^m=O(h_n^{-m}),
 \end{eqnarray}
 and $e^{C_5({\sf E}\left[|Y_i|^m\right])^{1/m}}\leq{\sf E}\left[e^{C_5|Y_i|}\right]\leq C_6$ by Assumption \ref{a6} and Jensen inequality.
 By Corollary 1.1 in \cite{Bo98} together with (\ref{eq6.9}), we have, for any $m>2$,
 \begin{eqnarray*}
 &&\left|{\sf Cov}\left(Y_iI_iK_{h_n}(X_i-x),Y_jI_jK_{h_n}(X_j-x)\right)\right|\notag\\
 &\leq&2m/(m-2) \cdot(h_n^{-m}\cdot{\sf E}[|Y_i|^m]\cdot\Vert K\Vert_\infty^m)^{2/m}\cdot[2\alpha([i-j])]^{1-2/m}\notag\\
 &\leq& c_{4,2}(m) \cdot h_n^{-2}\cdot[\alpha([i-j])]^{1-2/m}\notag
 \end{eqnarray*}
 with $c_{4,2}(m)=2^{2-2/m}\cdot m/(m-2)({\sf E}\left[|Y_i|^m\right]\|K\|_\infty^m)^{2/m}$.
 Let $m$ tend to infinity, we have $c_{4,2}(m)\to 4\cdot(\log(C_6)/C_5)^{2}\|K\|_\infty^2:=c_{4,2}$, thus
 \begin{equation*}\label{eq6.10}
 \left|{\sf Cov}\left(Y_iI_iK_{h_n}(X_i-x),Y_jI_jK_{h_n}(X_j-x)\right)\right|
 \leq c_{4,2} \cdot h_n^{-2}\cdot\alpha([i-j]).
 \end{equation*}
Using $\alpha(|i-j|)\leq C_1\rho^{|i-j|}$ in Assumption \ref{a2} or $\alpha(|i-j|)\leq C_1|i-j|^{-\gamma}$ in  Assumption \ref{a1},
and
taking $c_4=\max\{c_{4,1},c_{4,2}C_1\}$, we can prove (\ref{eq6.5}).\hfill$\Box$

\smallskip

The following lemma shows the uniform bound of the N-W estimator of a PSM process. Note that (i) the bandwidth is selected based on the sample size $n$, (ii) the estimator $\widehat f_{s,u}(x;h_n)$ is constructed based on subsample set $\{X_t,Y_t\}_{t=s}^u$, for $1\leq s\leq u\leq n$, and  (iii) the uniform bound is considered with respect to time $t$.

\begin{lemma}\label{lemma3}
Suppose the process $\{X_t,Y_t\}_{t=1}^n$ is PSM and Assumptions \ref{a3}--\ref{a6} are satisfied.
Let  $\widehat f_{1,t}(x;h_n)$ and $\widehat f_{t+1,n}(x;h_n)$ be defined in (\ref{eqf}). Then we have for $\forall x \in \mathbb{R}$, under the model (\ref{m1})
\begin{equation}\label{lemma31}
\max_{ \Delta_n\leq t\leq n-\Delta_n}\left|\widehat f_{1,t}(x;h_n)-{\sf E}\left[\widehat f_{1,t}(x;h_n)\right]\right|=O_{a.s.}\left(\frac{\log n}{\sqrt{nh_n}}\right),
\end{equation}
and
\begin{equation}\label{lemma32}
\max_{ \Delta_n\leq t\leq n-\Delta_n}\left|\widehat f_{t+1,n}(x;h_n)-{\sf E}\left[\widehat f_{t+1,n}(x;h_n)\right]\right|=O_{a.s.}\left(\frac{\log n}{\sqrt{nh_n}}\right).
\end{equation}
\end{lemma}
\smallskip

\noindent\textbf{Proof of Lemma \ref{lemma3}}

It is clear that if for some $\eta>0$,
\begin{equation}\label{eq6.13}
\sum_{n=1}^\infty{\sf P}\left(\max_{ \Delta_n\leq t\leq n-\Delta_n}\left|\widehat f_{1,t}(x;h_n)-{\sf E}\left[\widehat f_{1,t}(x;h_n)\right]\right|>\eta\cdot \frac{\log n}{\sqrt{nh_n}}\right)<\infty,
\end{equation}
we can show (\ref{lemma31}) by using the Borel-Cantelli lemma. Next we prove (\ref{eq6.13}). Actually,
\begin{eqnarray*}
&&{\sf P}\left(\max_{ \Delta_n\leq t\leq n-\Delta_n}\left|\widehat f_{1,t}(x;h_n)-{\sf E}\left[\widehat f_{1,t}(x;h_n)\right]\right|>\eta\cdot \frac{\log n}{\sqrt{nh_n}}\right)\notag\\
&\leq&\sum_{t=\Delta_n}^{n-\Delta_n}{\sf P}\left(\left|\widehat f_{1,t}(x;h_n)-{\sf E}\left[\widehat f_{1,t}(x;h_n)\right]\right|>\eta\cdot \frac{\log n}{\sqrt{nh_n}}\right)\notag\\
&\leq&\sum_{t=\Delta_n}^{n-\Delta_n}{\sf P}\left(\left|\widehat f_{1,t}(x;h_n)-{\sf E}\left[\widehat f_{1,t}(x;h_n)\right]\right|>\eta\sqrt{\delta}\cdot \frac{\log t}{\sqrt{th_n}}\right),\notag
\end{eqnarray*}
noting that $\log n/\sqrt{n}>\sqrt{\delta} \log t/\sqrt{t}$ when $\lfloor n\delta\rfloor\leq t<n$. A sufficient condition is that for some $\eta>0$ and $\delta_1>0$
\begin{equation}\label{eq6.14}
{\sf P}\left(\left|\widehat f_{1,t}(x;h_n)-{\sf E}\left[\widehat f_{1,t}(x;h_n)\right]\right|>\eta\cdot \frac{\log t}{\sqrt{th_n}}\right)
\leq c_5\cdot n^{-(2+\delta_1)}
\end{equation}
when $n$ is large enough, where $c_5$ is a positive constant (we still use the notation $\eta$ for $\eta\sqrt{\delta}$).

Next we prove (\ref{eq6.14}) using Lemma \ref{ppst:mixing}(ii).
Let $Z_{1,s,n}=K_{h_n}(x-X_s)-{\sf E}\left[K_{h_n}(x-X_s)\right]$ for $s=1,\cdots,n$, and denote the partial sum of $Z_{1,s,n}$ as $S_t=\sum _{s=1}^{t}Z_{1,s,n}$.

Firstly, we want to derive the order of $\sigma^2(q)$ and $v^2(q)$  defined in Lemma \ref{ppst:mixing} with the sequence $\{X_t\}_{t=1}^n$ replaced by the sequence $\{Z_{1,s,n}\}_{s=1}^t$. Taking
$\varepsilon=\varepsilon_t=(th_n)^{-1/2}\log t$, $q=q_t=\lfloor t^{1/2}h_n^{-1/2}\rfloor$, and $p=t/(2q)$, we have $|Z_{1,s,n}|\leq \Vert K \Vert_{\infty}\cdot h_n^{-1}$ and $q_t\leq t/2$ for large n.
Using the partition method similar to the proof of Theorem 3.3 in \cite{JR11}, we have (define $p'=\lfloor p\rfloor+2$)
\begin{eqnarray*}
\sigma^2(q)&\leq&\sum_{i=1}^{p'}{\sf Var}(K_{h_n}(X_i-x))+2\sum_{i>j}^{p'}|{\sf Cov}(K_{h_n}(X_i-x),K_{h_n}(X_j-x))|\notag\\
&=&\sum_{i=1}^{p'}{\sf Var}(K_{h_n}(X_i-x))+2\sum_{i=2}^{p'}(p'-i+1)|{\sf Cov}(K_{h_n}(X_i-x),K_{h_n}(X_1-x))|\notag\\
&\leq&c_1p'h_n^{-1}
+2p'\sum_{i=2}^B|{\sf Cov}(K_{h_n}(X_i-x),K_{h_n}(X_1-x))|\notag\\
&+&2p'\sum_{i=B+1}^{p'}|{\sf Cov}(K_{h_n}(X_i-x),K_{h_n}(X_1-x))|
\end{eqnarray*}
with the partition point $B=\lfloor h_n^{-q_F}\rfloor$.
Then using Lemma \ref{lemma2}, we can obtain for large n
\begin{eqnarray}\label{sigmaorder}
\sigma^2(q)&\leq&c_1p'h_n^{-1}+2p'\sum_{i=2}^B c_2h_n^{-1+q_F}+2p'\sum_{i=B+1}^{p'}c_2h_n^{-2}(i-1)^{-\gamma}\notag\\
&\leq&c_1p'h_n^{-1}+2p'B\cdot c_2 h_n^{-1+q_F}+2p'h_n^{-2}\cdot 2c_2 B^{1-\gamma}\notag\\
&\leq&c_1p'h_n^{-1}+2p'c_2h_n^{-q_F}\cdot h_n^{-1+q_F}+4c_2p'h_n^{-2+q_F(\gamma-1)}\notag\\
&\leq&(c_1+6c_2)\cdot p'h_n^{-1},
\end{eqnarray}
where the term $B^{1-\gamma}$ is induced by substituting the sum with an integral and the last row follows from $q_F(\gamma-1)>1$ in Assumption \ref{a5}. So for $v^2(q)$, we have
\begin{eqnarray}\label{vorder}
v^2(q)&=&2\sigma^2(q)/p^2+\Vert K \Vert_{\infty}\cdot h_n^{-1}\varepsilon_t/2\notag\\
&\leq&2(c_1+6c_2)h_n^{-1}p'p^{-2}+\Vert K \Vert_{\infty}\cdot h_n^{-1}\varepsilon_t/2\notag\\
&\leq& \Vert K \Vert_{\infty}\cdot h_n^{-1}\varepsilon_t
\end{eqnarray}
 for $n$ large enough, noting that $p'/(p^2)\simeq2\varepsilon_t/\log t=o(\varepsilon_t)$ when $\lfloor\delta n\rfloor\leq t\leq n-\lfloor\delta n\rfloor$. Then using Lemma \ref{ppst:mixing}(ii) and (\ref{vorder}),
 %together with the monotonicity of $\varepsilon_t$ with respect to $t$,
 we have for $\eta>0$
\begin{eqnarray}\label{eq6.15}
(6.13)&=&{\sf P}\left(\left|S_t\right|>t\cdot\eta \varepsilon_t\right)\notag\\
&\leq&4\exp\left(-\frac{\eta^2\varepsilon_t}{8 \Vert K\Vert_{\infty}}q_th_n\right)
+22\left(1+\frac{4\Vert K \Vert_{\infty}\cdot h_n^{-1}}{\eta\varepsilon_t}\right)^{1/2}q\alpha\left(\left\lfloor\frac{t}{2q_t}\right\rfloor\right)\notag\\
&:=&A_{1,t}+A_{2,t}.
\end{eqnarray}
Because $q_t\simeq t^{1/2}h_n^{-1/2}$ and thereby $\varepsilon_tq_th_n\simeq(\log t)$,
by selecting $\eta>\sqrt{8(2+\delta_1)\|K\|_\infty}$,
we have
$A_{1,t}\leq 4t^{-{\eta^2}/(8\|K\|_\infty)}=O\left(n^{-(2+\delta_1)}\right)$.
In terms of $A_{2,t}$, noting that $(h_n^{-1}/\varepsilon_t)^{1/2}q_t\leq t^{3/4}h_n^{-3/4}\leq n^{3/4}h_n^{-3/4}\rightarrow \infty$ and $\alpha\left(\left\lfloor\frac{t}{2q_t}\right\rfloor\right)
\leq C_1 {\left\lfloor\frac{t}{2q_t}\right\rfloor}^{-\gamma}\leq C_1 {\left\lfloor\frac{\sqrt{th_n}}{2}\right\rfloor}^{-\gamma}=O(n^{-\gamma/2}h_n^{-\gamma/2})$, we can obtain
\begin{eqnarray*}
A_{2,t}=O\left(n^{-\gamma/2+3/4}h_n^{-3/4-\gamma/2}\right)=O\left(n^{-(2+\delta_1)}\right)
\end{eqnarray*}
when $h_n\simeq n^{-\omega}$ for some $0<\omega\leq\frac{\gamma/2-11/4-\delta_1}{\gamma/2+3/4}<1-\frac{14}{2\gamma+3}$.

The proof of (\ref{lemma32}) is similar to the proof of (\ref{lemma31}) by considering the sequence $\{Z_{1,s,n}\}_{s=t+1}^n$. Thus we omit the proof.
Then we complete the proof of this lemma.\hfill$\Box$
\smallskip

\begin{lemma}\label{lemma4} Suppose the process $\{X_t,Y_t\}_{t=1}^n$ is GSM and Assumptions \ref{a3}--\ref{a6}  are satisfied.
Let  $\widehat f_{1,t}(x;h_n)$ and $\widehat f_{t+1,n}(x;h_n)$ be defined in (\ref{eqf}),
then we have  for $\forall x \in \mathbb{R}$, under the model (\ref{m1})
\begin{equation}\label{lemma3*1}
\max_{\ \Delta_n\leq t\leq n-\Delta_n}\left|\widehat f_{1,t}(x;h_n)-{\sf E}\left[\widehat f_{1,t}(x;h_n)\right]\right|=O_{a.s.}\left(\frac{\log n}{\sqrt{nh_n}}\right),
\end{equation}
and
\begin{equation}\label{lemma3*2}
\max_{\ \Delta_n\leq t\leq n-\Delta_n}\left|\widehat f_{t+1,n}(x;h_n)-{\sf E}\left[\widehat f_{t+1,n}(x;h_n)\right]\right|=O_{a.s.}\left(\frac{\log n}{\sqrt{nh_n}}\right).
\end{equation}
\end{lemma}

\noindent\textbf{ Proof of Lemma \ref{lemma4}}

The proof of this lemma is similar to the proof of Lemma \ref{lemma3}. Because of similarity, we only prove (\ref{lemma3*1}).

We need to prove (\ref{eq6.14}) by Lemma \ref{ppst:mixing}(ii). Using the same notation with Lemma \ref{lemma3}, we have $\sigma^2(q)=O(p'h_n^{-1})$, hence $v^2(q)\leq \Vert K \Vert_{\infty}\cdot h_n^{-1}\varepsilon_t$ for $n$ large enough (see Lemma 2.1 of \cite{Bo98}). Then we still have (\ref{eq6.15}). By selecting $\eta>\sqrt{8(2+\delta_1)\|K\|_\infty}$,
we have
\begin{equation}\label{eq6.19}
A_{1,t}\leq 4t^{-{\eta^2}/(8\|K\|_\infty)}=O\left(n^{-(2+\delta_1)}\right).
\end{equation}
In terms of $A_{2,t}$,  note that $\log t/\sqrt{th_n}\rightarrow0$, which implies that  $th_n$ and thus $\log t-\log h_n^{-1}\rightarrow\infty$, and therefore $\log t$ and $\log h_n^{-1}$ can be bounded by $\sqrt{th_n}$. We have
\begin{eqnarray}\label{eq6.20}
A_{2,t}&=&
22\left(1+\frac{4\Vert K \Vert_{\infty}\cdot h_n^{-1}}{\eta\varepsilon_t}\right)^{1/2}q\alpha\left(\left\lfloor\frac{t}{2q_t}\right\rfloor\right)\notag\\
&\leq& c_7\cdot t^{3/4}h_n^{-3/4}\cdot\rho^{\sqrt{th_n}/2}\notag\\
&=& c_7\cdot \exp\left\{\frac{3}{4}\left(\log t+\log h_n^{-1}\right)\right\}\cdot \exp\left\{-\frac{1}{2}\log \left(\frac{1}{\rho}\right)\cdot\sqrt{th_n}\right\}\notag\\
&\leq& c_7\cdot  \exp\left\{-c_8\sqrt{th_n}\right\}\notag\\
&=& c_7\cdot t^{ -c_8/\varepsilon_{t}} =o\left(n^{-(2+\delta_1)}\right),
\end{eqnarray}
where $c_7$ and $c_8$ are two positive constants, noting that $\varepsilon_t\rightarrow0$.
Combining (\ref{eq6.15}), (\ref{eq6.19}) and (\ref{eq6.20}), we can prove  (\ref{eq6.14}). Thus we complete the proof of this lemma.\hfill$\Box$

\smallskip

\begin{lemma}\label{lemma5}
Suppose the process $\{X_t,Y_t\}_{t=1}^n$ is PSM and Assumptions \ref{a3}--\ref{a6} are satisfied.
	Let $\widehat g_{1,t}(x;h_n)$ and $\widehat g_{t+1,n}(x;h_n)$ be defined in (\ref{eqg}). Then we have for $\forall x \in \mathbb{R}$, under the model (\ref{m1})
	\begin{equation}\
	\max_{\Delta_n\leq t\leq n- \Delta_n}\left|\widehat g_{1,t}(x;h_n)-{\sf E}[\widehat{g}_{1,t}(x;h_n)] \right |=O_{a.s.}\left(\frac{\log^2n}{\sqrt{nh_n}}\right),
	\end{equation}
and
	\begin{equation}\
	\max_{\Delta_n\leq t\leq n- \Delta_n}\left|\widehat g_{t+1,n}(x;h_n)-{\sf E}[\widehat{g}_{t+1,n}(x;h_n)]  \right|=O_{a.s.}\left(\frac{\log^2n}{\sqrt{nh_n}}\right),
	\end{equation}
\end{lemma}
\smallskip

\noindent\textbf{Proof of Lemma \ref{lemma5}}

 The proof of this lemma is similar to that of Lemma \ref{lemma3}. The only difference is that $Y_i$ may not be bounded, and we need to adopt the idea of truncation before using Lemma \ref{ppst:mixing}(ii).
Analogously, our goal is to prove for some $\eta>0$
\begin{eqnarray*}\label{gsum}
&&\sum_{n=1}^\infty{\sf P}\left(\max_{ \Delta_n\leq t\leq n-\Delta_n}\left|\widehat g_{1,t}(x;h_n)-{\sf E}\left[\widehat g_{1,t}(x;h_n)\right]\right|>\eta \frac{\log^2 n}{\sqrt{nh_n}}\right)<\infty,
\end{eqnarray*}
which can be proved by
\begin{equation}\label{eq6.24}
{\sf P}\left(\left|\widehat g_{1,t}(x;h_n)-{\sf E}[\widehat g_{1,t}(x;h_n)]\right|>\eta \varepsilon_t\right)=O\left(n^{-(2+\delta_1)}\right),
\end{equation}
where $\varepsilon_t=\log^2t/\sqrt{th_n}$ and $\delta_1$ is a tiny positive constant.

Next, we prove (\ref{eq6.24}). For $s=1,\cdots,n$ , define
$\bar{Y}_s=Y_sI_{(|Y_s|\leq T_{t})}$ and $\widetilde{Y}_s=Y_sI_{(|Y_s|>T_{t})}$   with $T_{t}=c_9\log t$, where $c_9$ is a positive constant which will be determined later. Then
\begin{eqnarray}\label{eq6.25}
Z_{2,s,n}&=:&Y_sK_{h_n}(X_s-x)-{\sf E}[Y_sK_{h_n}(X_s-x)]\notag\\
&=&(\bar{Y}_sK_{h_n}(X_s-x)-{\sf E}[\bar{Y}_sK_{h_n}(X_s-x)])\notag\\
&&+(\widetilde{Y}_sK_{h_n}(X_s-x)-{\sf E}[\widetilde{Y}_sK_{h_n}(X_s-x)])\notag\\
&:=&\bar{Z}_{2,s,n}+\widetilde{Z}_{2,s,n}.
\end{eqnarray}
Denote the partial sums in (\ref{eq6.25}) as $\bar S_{t,n}=\sum_{s=1}^t\bar{Z}_{2,s,n}$
and $\widetilde S_{t,n}=\sum_{s=1}^t\widetilde{Z}_{2,s,n}$. To use Lemma \ref{ppst:mixing}(ii), set $q=q_t=\lfloor t^{1/2}h_{n}^{-1/2}\rfloor$. Then (\ref{eq6.24}) can be written as follows,
\begin{eqnarray*}
{\sf P}\left(\left|\widehat g_{1,t}(x;h_n)-{\sf E}[\widehat g_{1,t}(x;h_n)]\right|>\eta \varepsilon_t\right)
&=&{\sf P}\left(|\bar S_{t,n}+\widetilde S_{t,n}|>t\cdot \eta \varepsilon_t\right)\notag\\
&\leq&{\sf P}\left(|\bar S_{t,n}|>t\cdot \frac{\eta}{2} \varepsilon_t\right)
+{\sf P}\left(|\widetilde S_{t,n}|>t \cdot\frac{\eta}{2} \varepsilon_t\right),\notag
\end{eqnarray*}
so we can prove this lemma by showing that
\begin{equation}\label{eq6.26}
{\sf P}\left(|\bar S_{t,n}|>t\cdot \frac{\eta}{2} \varepsilon_t\right)=O\left(n^{-(2+\delta_1)}\right)
\end{equation}
and
\begin{equation}\label{eq6.27}
{\sf P}\left(|\widetilde S_{t,n}|>t \cdot\frac{\eta}{2} \varepsilon_t\right)=O\left(n^{-(2+\delta_1)}\right).
\end{equation}

For  (\ref{eq6.26}), before using the similar method by the inequality in Lemma \ref{ppst:mixing}(ii) like before, we still need to show the bound of $\sigma^2(q)$.
Together with Lemma \ref{lemma2}(ii), it immediately follows that, like (\ref{sigmaorder}) by using
$B = \lfloor h_n^{-q_G}\rfloor$, for large n
\begin{eqnarray*}
\sigma^2(q)&\leq&c_3p'h_n^{-1}+2p'\sum_{i=2}^B c_4h_n^{-1+q_G}+2p'\sum_{i=B+1}^{p'}c_4h_n^{-2}(i-1)^{- \gamma}\notag\\
&\leq&c_3p'h_n^{-1}+2p'B\cdot c_4h_n^{-1+q_G}+2p'h_n^{-2}c_4\cdot2B^{1- \gamma}\notag\\
&\leq&c_3p'h_n^{-1}+2p'c_4h_n^{-q_G}\cdot h_n^{-1+q_G}+4p'c_4h_n^{-2+q_G( \gamma-1)}\notag\\
&\leq&(c_3+6c_4)\cdot p'h_n^{-1}\notag
\end{eqnarray*}
when $q_G( \gamma-1)>1$. Hence
\begin{eqnarray*}
v^2(q)\leq \left(\frac{4(c_3+6c_4) }{c_9\|K\|_\infty \log^3 t}+1\right)\Vert K \Vert_{\infty} h_{n}^{-1}\varepsilon_t T_{t}
\leq2\Vert K \Vert_{\infty} h_{n}^{-1}\varepsilon_t T_{t}
\end{eqnarray*}
when n is sufficiently large.
Then we can use Lemma \ref{ppst:mixing}(ii) like the proof before and derive that for $\eta>0$
\begin{eqnarray*}
{\sf P}\left(|\bar S_{t,n}|>t\cdot \frac{\eta}{2}\varepsilon_t\right)
&\leq& 4\exp\left(-\frac{\eta^2\varepsilon_t}{64 \Vert K \Vert_{\infty}T_{t} }qh_n\right)
+22\left(1+\frac{16\|K\|_\infty h_n^{-1}T_{t}}{\eta\varepsilon_t}\right)^{1/2}q\alpha\left(\left\lfloor\frac{t}{2q}\right\rfloor\right)\notag\\
&:=&A_{3,t}+A_{4,t}.
\end{eqnarray*}
For $A_{3,t}$, we have
\begin{eqnarray*}
A_{3,t}&\simeq&4\exp\left(-\frac{\eta^2}{64c_9\Vert K \Vert_{\infty}}\cdot \log t \right)=4 t^{-\eta^2/(64c_9\Vert K \Vert_{\infty})}
=O\left(n^{-(2+\delta_1)}\right)
\end{eqnarray*}
by selecting $\eta>8\sqrt{c_9(2+\delta_1)\Vert K \Vert_{\infty}}$.
For $A_{4,t}$, we have
\begin{eqnarray*}
A_{4,t}&=&22\left(1+\frac{16c_2\|K\|_\infty \sqrt{t}}{\eta\sqrt{h_n}\log t}\right)^{1/2}\frac{\sqrt{t}}{\sqrt{h_n}}\left(\frac{\sqrt{th_n}}{4}\right)^{-\gamma}\notag\\
&\leq& c_{10}\frac{t^{3/4-\gamma/2}}{h_n^{3/4+\gamma/2}}\frac{1}{\log^{1/2} t}\notag\\
&\leq&c_{11}\frac{1}{n^{\gamma/2-3/4}h_n^{3/4+\gamma/2}}\notag\\
&=&O\left(n^{-(2+\delta_1)}\right),
\end{eqnarray*}
where the existence of $\delta_1$ in the last equality follows from Assumption 3.
Then we have proved (\ref{eq6.26}).

In terms of (\ref{eq6.27}), using Cauchy-Schwarz and Markov inequality, we have
\begin{eqnarray*}
{\sf P}\left(|\widetilde S_{t,n}|>t\cdot \frac{\eta}{2}\varepsilon_t\right)
&\leq&\frac{2{\sf E}\left[|\widetilde S_{t,n}|\right]}{t \eta \varepsilon_t}\notag\\
&\leq&\frac{2{\sf E}\left[|Y_1K_h(X_1-x)I_{(|Y_1|> T_{t})}|\right]}{ \eta \varepsilon_t}\notag\\
&\leq&\frac{2\left\{\mathsf{E}\left[\left|Y_1K_h(X_1-x)\right|^2\right]\right\}^{1/2}\left\{\mathsf{P}\left(C_5\left|Y_1\right|>C_5T_{t}\right)\right\}^{1/2}}{\eta\varepsilon_t}\notag\\
&=&\frac{2\left\{\int{\sf E}[Y_1^2|X_1=u]K_{h_n}^2(u-x)f(u)du\right\}^{1/2} \left\{\mathsf{P}(e^{C_5|Y_1|}>e^{C_5T_{t}})\right\}^{1/2}}{\eta\varepsilon_t}\notag\\
&=&O\left(\frac{\sqrt{th_n}}{\log^2 t}\cdot h_n^{-1/2}\cdot t^{-\frac{c_9C_5}{2}}\right)\notag\\
&=&O\left(n^{-(2+\delta_1)}\right)
\end{eqnarray*}
by letting $c_9>(5+2\delta_1)/C_5$, noting that $\|K\|_2<\infty$, $f$ is uniformly bounded, $\sup \limits_{x\in {\sf supp}[{ X_1}]}{\sf E}\left[Y_1^2|X_1=x\right]<\infty$, and ${\sf E}\left[e^{C_5|Y_1|}\right]<\infty$.

Combing (\ref{eq6.26}) and (\ref{eq6.27}), we  obtain (\ref{eq6.24}).
Hence the proof is completed.\hfill$\Box$

\smallskip

\begin{lemma}\label{lemma6}Suppose the process $(X_t,Y_t)$ is GSM and Assumptions \ref{a3}--\ref{a6} are satisfied.
Let $\widehat g_{1,t}(x;h_n)$ and $\widehat g_{t+1,n}(x;h_n)$ be defined in (\ref{eqg}). Then we have for $\forall x \in \mathbb{R}$, under the model (\ref{m1})
\begin{equation}
\max_{\Delta_n\leq t\leq n- \Delta_n}|\widehat g_{1,t}(x;h_n)-{\sf E}[\widehat{g}_{1,t}(x;h_n)]  |=O_{a.s.}\left(\frac{\log^2n}{\sqrt{nh_n}}\right),
\end{equation}
and
\begin{equation}
\max_{\Delta_n\leq t\leq n- \Delta_n}|\widehat g_{t+1,n}(x;h_n)-{\sf E}[\widehat{g}_{t+1,n}(x;h_n)]  |=O_{a.s.}\left(\frac{\log^2n}{\sqrt{nh_n}}\right).
\end{equation}
\end{lemma}

\noindent\textbf{Proof of Lemma \ref{lemma6}}

We still use the notation in Lemma \ref{lemma5}, and want to show (\ref{eq6.26}) and (\ref{eq6.27}). Together with Lemma \ref{lemma2}(ii), it still holds that
\begin{eqnarray*}\label{gsmysigma}
\sigma^2(q)&\leq& c_3p'h_n^{-1}+2p'\left(\sum_{i=2}^B c_4h_n^{-1+q_G}+\sum_{i=B+1}^{p'}c_4h_n^{-2}\rho^{(i-1) }\right)\notag\\
&\leq& (c_3+2c_4)p'h_n^{-1}+2c_4p' h_n^{-2}\cdot(1-\rho)^{-1}\rho^{h_n^{-q_G} }\notag\\
&=&O(p'h_n^{-1}),
\end{eqnarray*}
noting that $\frac{p'h_n^{-2}}{p'h_n^{-1}}\cdot\rho^{h_n^{-q_G} }=h_n^{-1}\rho^{h_n^{-q_G} }\simeq (h_n^{-q_G} )^{\frac{1}{q_G}}\cdot\rho^{h_n^{-q_G} }\rightarrow0$. So we only need to show the part $A_{4,t}$ containing mixing-coefficient as follows,
\begin{eqnarray}
A_{4,t}&=&22\left(1+\frac{16\|K\|_\infty h_n^{-1}T_{t}}{\eta\varepsilon_t}\right)^{1/2}q\alpha\left(\left\lfloor\frac{t}{2q}\right\rfloor\right)
\simeq 22\left(1+\frac{16c_9\|K\|_\infty \sqrt{t}}{\eta\sqrt{h_n}\log t}\right)^{1/2}\frac{\sqrt{t}}{\sqrt{h_n}}\cdot\rho^{\sqrt{th_n}/2}\notag\\
&\leq&c_{12}\frac{t^{3/4}}{h_n^{3/4}}\frac{1}{\log^{1/2} t}\cdot\rho^{\sqrt{th_n}/2}\notag\\
&\leq&c_{13}\exp\{-c_{14}\log^2 t(\varepsilon_t)^{-1}\}\notag\\
&=&O\left(n^{-(2+\delta_1)}\right),
\end{eqnarray}
where $c_{12}, c_{13}$ and $c_{14}$ are strictly positive constants, noting that the last second row is deduced like (\ref{eq6.20}), and $\varepsilon_t\rightarrow 0$.
\hfill$\Box$
\smallskip

\begin{lemma}\label{lemma7}Suppose that the assumptions in Lemmas \ref{lemma3} and \ref{lemma5} (or Lemmas \ref{lemma4} and \ref{lemma6}) are satisfied. Let  $\widehat \varphi_{1,t}(x;h_n)$ and $\widehat \varphi_{t+1,n}(x;h_n)$ be defined in (\ref{eqvarphi}). If the grid point $x_i\in \mathcal{X}$, then we have under the model (\ref{m1})
\begin{eqnarray}
\max \limits_{\Delta_n\leq t \leq n-\Delta_n}\left|\widehat{\varphi}_{1,t}(x_i;h_n)-\frac{{\sf E}[\widehat{g}_{1,t}(x_i;h_n)]}{{\sf E}[\widehat{f}_{1,t}(x_i;h_n)]}\right|=O_{a.s.}\left(\frac{\log^2 n}{\sqrt{nh_n}}\right),
\end{eqnarray}
and
\begin{eqnarray}
\max \limits_{\Delta_n\leq t \leq n-\Delta_n}\left|\widehat{\varphi}_{t+1,n}(x_i;h_n)-\frac{{\sf E}[\widehat{g}_{t+1,n}(x_i;h_n)]}{{\sf E}[\widehat{f}_{t+1,n}(x_i;h_n)]}\right|=O_{a.s.}\left(\frac{\log^2 n}{\sqrt{nh_n}}\right).
\end{eqnarray}
\end{lemma}

\noindent\textbf{Proof of Lemma \ref{lemma7}}

Because of similarity, we only show the first equation.
Consider the decomposition
\begin{eqnarray*}
&&\widehat{\varphi}_{1,t}(x_i)-\frac{{\sf E}[\widehat{g}_{1,t}(x_i;h_n)]}{{\sf E}[\widehat{f}_{1,t}(x_i;h_n)]}\notag\\
&=& \frac{\widehat{g}_{1,t}(x_i;h_n)-\widehat{f}_{1,t}(x_i;h_n)\frac{{\sf E}[\widehat{g}_{1,t}(x_i;h_n)]}{{\sf E}[\widehat{f}_{1,t}(x_i;h_n)]}}{{\sf E}[\widehat{f}_{1,t}(x_i;h_n)]}\notag\\
&&-\frac{\widehat{f}_{1,t}(x_i;h_n)-{\sf E}[\widehat{f}_{1,t}(x_i;h_n)]}{\widehat{f}_{1,t}(x_i;h_n)}\cdot
\frac{\widehat{g}_{1,t}(x_i;h_n)-\widehat{f}_{1,t}(x_i;h_n)\frac{{\sf E}[\widehat{g}_{1,t}(x_i;h_n)]}{{\sf E}[\widehat{f}_{1,t}(x_i;h_n)]}}{{\sf E}[\widehat{f}_{1,t}(x_i;h_n)]}.
\end{eqnarray*}

By Lemma 3, $\widehat{f}_{1,t}(x_i;h_n)-{\sf E}[\widehat{f}_{1,t}(x_i;h_n)]=o_{a.s.}(1)$
 uniformly over $t$. Since
 $x_i\in{\cal X}=({\sf supp} [X_1])^\circ$,
  the density function $f$ has a nonzero lower bound in a sufficient small neighbourhood of $x_i$. We have
  ${\sf E}[\widehat{f}_{1,t}(x_i;h_n)]={\sf E}\left[K_{h_n}({X_1-x_i})\right]=h_n^{-1}\int_{\mathbb{R}} K((u-x_i)/h_n)f(u)du=\int_{\mathbb{R}} K(z)f(z h_n+x_i)dz\geq c_{15}$  for some positive constant $c_{15}$ if $h_n$ is small enough.
 Then, by Lemma \ref{lemma3} and \ref{lemma5}, we have
\begin{eqnarray*}
&&\max \limits_{\Delta_n\leq t \leq n-\Delta_n}\left|\widehat{\varphi}_{1,t}(x_i;h_n)-\frac{{\sf E}[\widehat{g}_{1,t}(x_i;h_n)]}{{\sf E}[\widehat{f}_{1,t}(x_i;h_n)]}\right|\notag\\
&=&\max \limits_{\Delta_n\leq t \leq n-\Delta_n}\left|\frac{\widehat{g}_{1,t}(x_i;h_n)-\widehat{f}_{1,t}(x_i;h_n)\frac{{\sf E}[\widehat{g}_{1,t}(x_i;h_n)]}{{\sf E}[\widehat{f}_{1,t}(x_i;h_n)]}}{{\sf E}[\widehat{f}_{1,t}(x_i;h_n)]}\right|\left(1+o_{a.s.}(1)\right)\notag\\
&\leq&\max \limits_{\Delta_n\leq t \leq n-\Delta_n}\left| \frac{\widehat{g}_{1,t}(x_i;h_n)-{\sf E}[\widehat{g}_{1,t}(x_i;h_n)]}{{\sf E}[\widehat{f}_{1,t}(x_i;h_n)]}\right|\left(1+o_{a.s.}(1)\right)\notag\\
&&+\max \limits_{\Delta_n\leq t \leq n-\Delta_n}\left| \frac{\left(\widehat{f}_{1,t}(x_i;h_n)-{\sf E}[\widehat{f}_{1,t}(x_i;h_n)]\right)\frac{{\sf E}[\widehat{g}_{1,t}(x_i;h_n)]}{{\sf E}[\widehat{f}_{1,t}(x_i;h_n)]}}{{\sf E}[\widehat{f}_{1,t}(x_i;h_n)]}\right|\left(1+o_{a.s.}(1)\right)\notag\\
&=&O_{a.s.}\left(\frac{\log^2 n}{\sqrt{nh_n}}\right),
\end{eqnarray*}
noting that ${\sf E}\left[|\widehat{g}_{1,t}(x_i;h_n)|\right]\leq{\sf E}\left[|Y_iK_{h_n}(X_1-x_i)|\right]=\int_{\mathbb{R}}{\sf E}\left[|Y_1|\mid X_1=h_nz+x_i\right]\cdot K(z)f(h_nz+x_i)dz\leq C_4||K||_1||f||_\infty$.

Hence we complete the proof of this lemma.
\hfill$\Box$
\bigskip

\noindent{ \textbf{Proof of Theorem \ref{theorem1}}}

Since there is no change point, we have
${\sf E}[\widehat{f}_{1,t}(x_i;h_n)]={\sf E}[\widehat{f}_{t+1,n}(x_i;h_n)]$
and
${\sf E}[\widehat{g}_{1,t}(x_i;h_n)]={\sf E}[\widehat{g}_{t+1,n}(x_i;h_n)]$,
and therefore by Lemma \ref{lemma7}
\begin{eqnarray*}
\max \limits_{\Delta_n\leq t \leq n-\Delta_n}W_{1,n}(t)&\leq&2\max \limits_{\Delta_n\leq t \leq n-\Delta_n}\sum_{i=1}^{m}\left|\widehat{\varphi}_{1,t}(x_i;h_n)-\frac{{\sf E}[\widehat{g}_{1,t}(x_i;h_n)]}{{\sf E}[\widehat{f}_{1,t}(x_i;h_n)]}\right|^2\notag\\
&+&2\max \limits_{\Delta_n\leq t \leq n-\Delta_n}\sum_{i=1}^{m}\left|\widehat{\varphi}_{t+1,n}(x_i;h_n)-\frac{{\sf E}[\widehat{g}_{t+1,n}(x_i;h_n)]}{{\sf E}[\widehat{f}_{t+1,n}(x_i;h_n)]}\right|^2\notag\\
&=&O_{a.s.}\left(\frac{\log^4 n}{nh_n}\right).
\end{eqnarray*}
We complete the proof of the theorem.
\hfill$\Box$

\bigskip

\noindent{\textbf{Proof of Theorem \ref{theorem2}}}

By definition,
\begin{eqnarray*}
\Lambda_{h_n}^2(x_i)&=&\left({\sf E}\left[\left(\varphi_1(X_1)-\varphi_2(X_1)\right)K_{h_n}(x_i-X_1)\right]\right)^2/({\sf E}[K_{h_n}(X_1-x_i)])^2\\
&=&\left(\int_{\mathbb{R}}(\varphi_1(u)-\varphi_2(u))K_{h_n}(x_i-u)f(u)du\right)^2/({\sf E}[K_{h_n}(X_1-x_i)])^2\notag\\
&=&\left(\int_{u\in\mathcal{X}\cap\mathcal{Y}}\left(\varphi_1(u)-\varphi_2(u)\right)K_{h_n}(x_i-u)f(u)du\right)^2/({\sf E}[K_{h_n}(X_1-x_i)])^2\notag\\
&=&\left(\int_{(h_nz+x_i)\in\mathcal{X}\cap\mathcal{Y}}\left(\varphi_1(h_nz+x_i)-\varphi_2(h_nz+x_i)\right)K(z)f(h_nz+x_i)dz\right)^2/({\sf E}[K_{h_n}(X_1-x_i)])^2.\notag
\end{eqnarray*}
Note that $({\sf E}[K_{h_n}(X_1-x_i)])$ is bounded. When $x_i\in \mathcal{X}\cap\mathcal{Y}$,
both $\varphi_1(x)-\varphi_2(x)$ and $f(x)$ are bounded away from zero in a small neighbour of $x_i$.
Therefore $\Lambda_{h_n}^2(x_i)$ is also bounded away from zero when $n$ is large enough.

Next we prove (\ref{eqT2}).
 Considering the special point $t=k$, it is obvious that
\begin{eqnarray*}
\max \limits_{\Delta_n\leq t< n-\Delta_n}W_{1,n}(t)\geq W_{1,n}(k).
\end{eqnarray*}
Note that  the sequences
$\{ X_s,Y_s \}_{s=1}^k$ and $\{ X_s,Y_s \}_{s=k+1}^n$ are strictly stationary.
By definition and Lemma \ref{lemma7}, we have
\begin{eqnarray}
W_{1,n}(k)&=&\frac{k(n-k)}{n^2}\sum_{i=1}^m\left|\widehat \varphi_{1,k}(x_i;h_n)-\widehat \varphi_{k+1,n}(x_i;h_n)\right|^2\notag\\
&=&\theta(1-\theta)\sum_{i=1}^m\left(\frac{{\sf E}[\widehat{g}_{1,k}(x_i;h_n)]}{{\sf E}[\widehat{f}_{1,k}(x_i;h_n)]}-\frac{{\sf E}[\widehat{g}_{k+1,n}(x_i;h_n)]}{{\sf E}[\widehat{f}_{k+1,n}(x_i;h_n)]}\right)^2+O_{a.s.}\left(\frac{\log^2 n}{\sqrt{nh_n}}\right)\notag\\
&=&\theta(1-\theta)\sum_{i=1}^m\Lambda^2_{h_n}(x_i)+O_{a.s.}\left(\frac{\log^2 n}{\sqrt{nh_n}}\right).
\end{eqnarray}

Thus, we complete the proof of this theorem.
\hfill$\Box$

\medskip

\noindent{\textbf{Proof of Theorem \ref{theorem3}}}

(i) It is a direct corollary from Theorem \ref{theorem1} and \ref{theorem2}.

Now we prove (ii). From the statement in the proof of Theorem \ref{theorem2}, we have
\begin{eqnarray}\label{3.3max1}
\max \limits_{\Delta_n\leq t\leq k}W_{1,n}(t)\geq\theta(1-\theta)\sum_{i=1}^m\Lambda^2_{h_n}(x_i)
\end{eqnarray}
almost surely when $n\rightarrow\infty$. If we can show that
\begin{eqnarray}\label{3.3max2}
\max \limits_{\Delta_n\leq t\leq k-n\varepsilon}W_{1,n}(t)<\theta(1-\theta)\sum_{i=1}^m\Lambda^2_{h_n}(x_i)
\end{eqnarray}
almost surely, for any small $\epsilon>0$ when $n\rightarrow\infty$,
then (\ref{3.3max1}) and (\ref{3.3max2}) imply that $\widehat k\geq k-n\varepsilon$ almost surely when $n\rightarrow\infty$. Using the same method for the case when $k+n\varepsilon\leq t\leq n-\Delta_n$, we can obtain $\widehat k\leq k+n\varepsilon$ almost surely when $n\rightarrow\infty$. Combining these two inequalities we can show that $(\widehat k-k)/n=o_{a.s.}(1)$ by letting $\varepsilon\rightarrow0$.

Next we prove (\ref{3.3max2}). %As the proof of (\ref{W(k)equality}), we consider the medium
%${\sf E}[\widehat{g}_{1,t}(x_i;h_n)]/{\sf E}[\widehat{f}_{1,t}(x_i;h_n)]$ in the following proof.
When $t\leq k-n\varepsilon$, on the one hand, we have
\begin{eqnarray}\label{phi1}
\widehat{\varphi}_{1,t}(x_i;h_n)-\frac{{\sf E}[\widehat{g}_{1,t}(x_i;h_n)]}{{\sf E}[\widehat{f}_{1,t}(x_i;h_n)]}=O_{a.s.}\left(\frac{\log^2 n}{\sqrt{nh_n}}\right)
\end{eqnarray}
 uniformly in $t$ by Lemma \ref{lemma7}.
On the other hand,
\begin{eqnarray*}
&&\widehat \varphi_{t+1,n}(x_i;h_n)-\frac{{\sf E}[\widehat{g}_{1,t}(x_i;h_n)]}{{\sf E}[\widehat{f}_{1,t}(x_i;h_n)]}
=\frac{\widehat g_{t+1,n}(x_i;h_n)}{\widehat f_{t+1,n}(x_i;h_n)}-\frac{{\sf E}[\widehat{g}_{1,t}(x_i;h_n)]}{{\sf E}[\widehat{f}_{1,t}(x_i;h_n)]}\notag\\
&=&\frac{(k-t)\left(\widehat g_{t+1,k}(x_i;h_n)-{\sf E}[\widehat g_{t+1,k}(x_i;h_n)]\right)+(n-k)\left(\widehat g_{k+1,n}(x_i;h_n)-{\sf E}[\widehat g_{k+1,n}(x_i;h_n)]\right)}
{(n-t)\widehat f_{t+1,n}(x_i;h_n)}\notag\\
&+&\frac{(k-t){\sf E}[g_{t+1,k}(x_i;h_n)]+(n-k){\sf E}[g_{k+1,n}(x_i;h_n)]-(n-t)\widehat f_{t+1,n}(x_i;h_n)\frac{{\sf E}[\widehat{g}_{1,t}(x_i;h_n)]}{{\sf E}[\widehat{f}_{1,t}(x_i;h_n)]}}
{(n-t)\widehat f_{t+1,n}(x_i;h_n)}\notag\\
&:=&B_3(x_i)+B_4(x_i).
\end{eqnarray*}
We show that $B_3(x_i)$ is negligible and $B_4(x_i)$ is the leading term.
Similar to the argument of proof for Lemma 5, we can show that the asymptotic order of $\widehat g_{t+1,k}(x_i;h_n)-{\sf E}[\widehat g_{t+1,k}(x_i;h_n)]$
is also $\frac{\log^2 n}{\sqrt{nh_n}}$, as the sample size $k-t$ is of order $n$.
Note that  $\widehat f_{t+1,n}$  is bounded  away from zero almost surely when $n$ is large enough.
Since $(k-t)/(n-t)<\theta/(1-\theta)$ and $(n-k)/(n-t)<1$, we have
by Lemma 5 (or Lemma 6)
\begin{equation}\label{B3}
B_3(x_i)=O_{a.s.}\left(\frac{\log^2 n}{\sqrt{nh_n}}\right).
\end{equation}

In terms of $B_4(x_i)$, we have by Lemma 3
\begin{equation*}
(n-t)\widehat f_{t+1,n}(x_i;h_n)=(k-t){\sf E}[\widehat{f}_{1,k}(x_i;h_n)]+(n-k){\sf E}[\widehat{f}_{k+1,n}(x_i;h_n)]+O_{a.s.}\left(\frac{n\log n}{\sqrt{nh_n}}\right).
\end{equation*}
Therefore,
\begin{eqnarray}\label{B4}
B_4(x_i)&=&\frac
{(k-t){\sf E}[\widehat{g}_{1,k}(x_i;h_n)]+(n-k){\sf E}[\widehat{g}_{k+1,n}(x_i;h_n)]}
{(k-t){\sf E}[\widehat{f}_{1,k}(x_i;h_n)]+(n-k){\sf E}[\widehat{f}_{k+1,n}(x_i;h_n)]}
-\frac{{\sf E}[\widehat{g}_{1,t}(x_i;h_n)]}{{\sf E}[\widehat{f}_{1,t}(x_i;h_n)]}
+O_{a.s.}\left(\frac{\log^2 n}{\sqrt{nh_n}}\right)\notag\\
&=&\frac
{(k-t){\sf E}[\widehat{g}_{1,k}(x_i;h_n)]+(n-k){\sf E}[\widehat{g}_{k+1,n}(x_i;h_n)]}
{(k-t){\sf E}[\widehat{f}_{1,k}(x_i;h_n)]+(n-k){\sf E}[\widehat{f}_{k+1,n}(x_i;h_n)]}\notag\\
&&-\frac
{\left((k-t){\sf E}[\widehat{f}_{1,k}(x_i;h_n)]+(n-k){\sf E}[\widehat{f}_{k+1,n}(x_i;h_n)]\right)\frac{{\sf E}[\widehat{g}_{1,t}(x_i;h_n)]}{{\sf E}[\widehat{f}_{1,t}(x_i;h_n)]}}
{(k-t){\sf E}[\widehat{f}_{1,k}(x_i;h_n)]+(n-k){\sf E}[\widehat{f}_{k+1,n}(x_i;h_n)]}
+O_{a.s.}\left(\frac{\log^2 n}{\sqrt{nh_n}}\right)\notag\\
&=&\frac
{(n-k){\sf E}[\widehat{f}_{k+1,n}(x_i;h_n)]}
{(k-t){\sf E}[\widehat{f}_{1,k}(x_i;h_n)]+(n-k){\sf E}[\widehat{f}_{k+1,n}(x_i;h_n)]}
\cdot\left(
\frac{{\sf E}[\widehat{g}_{k+1,n}(x_i;h_n)]}{{\sf E}[\widehat{f}_{k+1,n}(x_i;h_n)]}-
\frac{{\sf E}[\widehat{g}_{1,t}(x_i;h_n)]}{{\sf E}[\widehat{f}_{1,t}(x_i;h_n)]}
\right)\notag\\
&&+O_{a.s.}\left(\frac{\log^2 n}{\sqrt{nh_n}}\right).
\end{eqnarray}

Combining (\ref{phi1})--(\ref{B4}), we can obtain uniformly in $t$
\begin{eqnarray*}
W_{1,n}(t)&=&\frac{t(n-t)}{n^2}\sum \limits_{i=1} \limits^{m}|\widehat \varphi_{1,t}(x_i;h_n)- \widehat \varphi_{t+1,n}(x_i;h_n)|^2\notag\\
&=&\frac{t(n-t)}{n^2}\sum \limits_{i=1} \limits^{m}B_4^2(x_i)+O_{a.s.}\left(\frac{\log^2 n}{\sqrt{nh_n}}\right)\notag\\
&=&\frac{t(n-t)}{n^2}\cdot\frac{(n-k)^2}{(n-t)^2}\sum \limits_{i=1} \limits^{m}\Lambda^2_{h_n}(x_i)+O_{a.s.}\left(\frac{\log^2 n}{\sqrt{nh_n}}\right)\notag\\
&=&\frac{t(n-k)^2}{n^2(n-t)}\sum \limits_{i=1} \limits^{m}\Lambda^2_{h_n}(x_i)+O_{a.s.}\left(\frac{\log^2 n}{\sqrt{nh_n}}\right).
\end{eqnarray*}
Note that 
$$\frac{t(n-k)^2}{n^2(n-t)}\leq(1-\theta)^2\frac{\theta-\varepsilon}{1-\theta+\varepsilon}
=(1-\theta)\cdot\frac{1-\theta}{1-\theta+\varepsilon}\cdot(\theta-\varepsilon)<\theta(1-\theta).$$
Thus, we have (\ref{3.3max2}), and we complete the proof.
\hfill$\Box$
\medskip

\end{document}